# Topological Phase Transitions and Berry-Phase Hysteresis in Exchange-Coupled Nanomagnets

Ahsan Ullah,[*] Xin Li, Yunlong Jin, Rabindra Pahari, Lanping Yue, Xiaoshan Xu, Balamurugan Balasubramanian, David J. Sellmyer,[‡] and Ralph Skomski[‡]

*Department of Physics & Astronomy and Nebraska Center for Materials and Nanoscience, University of Nebraska, Lincoln, NE 68588*

*E-mail: aullah@huskers.unl.edu
[‡]Deceased

**Abstract**

Topological phase in magnetic materials yields a quantized contribution to the Hall effect known as the topological Hall effect (THE), which is often caused by skyrmions, with each skyrmion creating a magnetic flux quantum $\pm h/e$. The control and understanding of topological properties in nanostructured materials is the subject of immense interest for both fundamental science and technological applications, especially in spintronics. In this article, the electron-transport properties and spin structure of exchange-coupled Co nanoparticles with an average particle size of 13.7 nm are studied experimentally and theoretically. Magnetic and Hall effect measurements identify topological phase transitions in the exchange-coupled Co nanoparticles and discover a qualitatively new type of hysteresis in the topological Hall effect, namely Berry-phase hysteresis. Micromagnetic simulations reveal the origin of the topological Hall effect, namely the chiral domains with domain-wall chirality quantified by an integer skyrmion number. These spin structures are different from the skyrmions formed due to Dzyaloshinskii-Moriya interactions in B20 crystals and multilayered thin films and caused by cooperative magnetization reversal in the exchange-coupled Co nanoparticles. An analytical model is developed to explain the underlying physics of the Berry-phase hysteresis, which is strikingly different from the iconic magnetic hysteresis and constitutes one aspect of 21$^{st}$ century reshaping of our view on nature at the borderline of physics, chemistry, mathematics, and materials science.

## I.  INTRODUCTION

Topological phase transitions (TPTs) permeate areas such as superfluid and superconductors [1, 2], basic quantum mechanics [3, 4], fractional quantum-Hall effect [5], and topological insulators [6] and therefore have gained significant interest in both science and technology. TPTs are very different from ordinary Landau-type phase transitions [4, 7-9]. Rather than involving symmetry breaking and order-parameter changes, they are characterized by changes in topological numbers. For example, coffee cups have one hole, located in the handle, and are therefore characterized by the topological number (Euler genus) $g = 1$. A flat pancake has no holes ($g = 0$) so that the piercing of a number of holes into a pancake is a trivial example of a TPT.

Topological phase transition is in contrast to magnetic hysteresis, which is based on a phase transition between an ordered low-temperature and a disordered high-temperature [1-6, 10, 11, 12]. An intriguing aspect of magnetic hysteresis is its relation to magnetic phase transitions. Figures 1(a-b) compare the atomic-scale origin of ferromagnetism with the nanoscale or 'micromagnetic' origin of hysteresis. When a ferromagnet is cooled below the Curie temperature $T_c$, it develops a spontaneous magnetization $M_s$ (a). This process is a Landau-type phase transition, defined as a singular change of a local order parameter ($M$) due to spontaneous symmetry breaking. The ordered phase has the character of a $k = 0$ Goldstone mode whose magnetization can point in any direction (a). This degeneracy is removed by symmetry-*violating* terms in the Hamiltonian, such as magnetic anisotropy [13].

Magnetic hysteresis, Fig. 1(b), is on top of the Landau transition (a). When a magnetic material is subjected to an external field $H$, then its magnetization $M(H)$ is generally not single-valued but splits into ascending and descending branches. A well-known example is small nanoparticles of volume $V$ and anisotropy energy $K_1 V \sin^2\theta$ in a magnetic field $H = H_z$. The color coding throughout this article is $M_z(r) = +M_s$ (red), $M_z(r) = -M_s$ (blue) and intermediate (yellow). For positive $K_1$, $\theta = 0$ (red) and $\theta = 180°$ (blue) are energetically favorable but separated by an energy barrier $K_1 V$ ($\theta = 90°$). This energy barrier needs some external field to be overcome and is therefore the reason for the hysteresis.

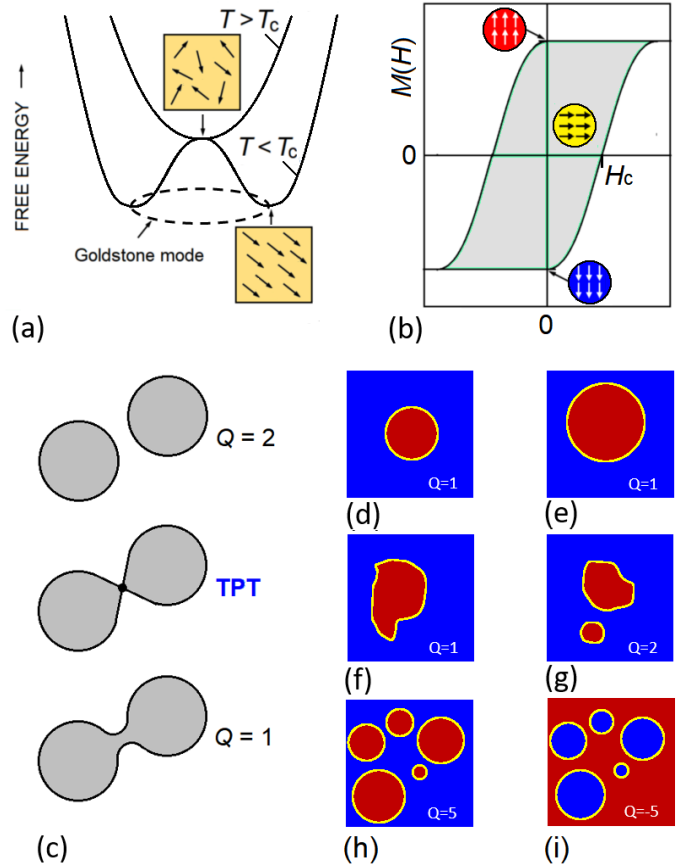

**Fig. 1. Phase transitions:** (a) Curie transition (magnetic Landau transition), (b) magnetic hysteresis, (c) Lifshitz transition in metal, and (d-i) topological phase transition in a magnetic thin film with perpendicular anisotropy. In (c), the gray areas denote the $k$-space region occupied by electrons at the Fermi level. In (d-i), red and blue regions indicate positive (↑) and negative (↓) magnetizations with respect to the film plane. Topological phase transitions are characterized by topological numbers $Q$. The topological protection in the micromagnetic case is experimentally established, for example through the "blowing" of skyrmions [14 - 16].

While topology has a long history, the idea of topological phase transition goes back to the Lifshitz transition [4, 7]. Figure 1(c) shows the $k$-space meaning of the Lifshitz transition in metals. Itinerant electrons fill the available electron states until the Fermi level is reached. The occupancy at the Fermi level (gray) depends on the number of electrons, and there are several scenarios that change the topological quantum number $Q$, such as external mechanical pressure and chemical addition of electrons. Each Fermi-surface region (gray) yields an integer contribution to $Q$, irrespective of the size and shape of the pocket. Topological concepts are now applied to many areas of physics, from skyrmions [7, 8, 13, 15, 16, 17- 20] to topological insulators and other quantum materials [21-30], all of them fascinating research topics in their own rights.

Figures 1(d-i) show the magnetic analog of the Lifshitz transition in a thin film. The field $H$ is perpendicular to the film and affects ↑ (red) and ↓ (blue) regions separated by domain walls (yellow). The underlying micromagnetism is very similar to that of magnetic skyrmions [31, 32, 33-36] and to XY-model transitions [8]. When an electrical current flows through the film, then the spins of the conduction electrons exchange-interact with the local magnetization $M(r)$ and become, in general, noncoplanar noncollinear. This noncollinearity creates a Berry curvature [5], an emergent magnetic field, and subsequently a Hall-effect contribution known as the topological Hall effect (THE) [5, 33]. These effects are proportional to the *skyrmion density* [31, 33, 37, 38]

$$\Phi = \frac{1}{4\pi} \boldsymbol{m} \cdot \left( \frac{\partial \boldsymbol{m}}{\partial x} \times \frac{\partial \boldsymbol{m}}{\partial y} \right) \qquad (1)$$

where $\boldsymbol{m} = \boldsymbol{M}(r)/M_s$ is the normalized magnetization and the $x$-$y$-plane is the film plane. The emergent magnetic flux that corresponds to the THE is equal to $Q\, h/e$, where $Q = \int \Phi\, dx dy$ is the skyrmion number and $h/e$ is the magnetic flux quantum. In granular thin films, such as the one considered in the present paper, there are also nonzero derivatives $\partial \boldsymbol{m}/\partial z$. By virtue of measurement geometry, $\partial \boldsymbol{m}/\partial z$ does not contribute to the THE [33], but it is one source of noise [39]. The skyrmion density is nonzero for spins $\boldsymbol{m}(\mathbf{r})$ that are both noncollinear and noncoplanar, and Eq. (1) is actually a continuum version of the triple product or spin chirality $\chi_s = \boldsymbol{m}_i \cdot (\boldsymbol{m}_j \times \boldsymbol{m}_k)$, where $\boldsymbol{m}_i = \boldsymbol{m}(\boldsymbol{R}_i)$ describes the atomic spins that cause the conduction electrons to develop their Berry phase.

Much of the fascination with topological phase transitions originate from the great simplicity of the mathematics conveyed by Eq. (1), which is summarized in Supplement A [45]. In skyrmionic structures such as those of Figs. 1(d-i), the spins inside the red and blue regions are parallel ($m = \pm e_z$), so that $\partial \boldsymbol{m}/\partial x$, $\partial \boldsymbol{m}/\partial y$, and $\Phi$ are zero. The integral over $\Phi$ therefore reduces to an integral over the yellow domain-boundary regions in Figs. 1(d-i). It can be shown that

$$Q = \frac{1}{2\pi} \oint \boldsymbol{\kappa} \cdot d\boldsymbol{l} \qquad (2)$$

where $\kappa$ is the curvature of the region's yellow boundary and the integral in Eq. (7) has the value $2\pi$ [40]. This integral is equal to $\pm 1$ for any area enclosed by a single yellow boundary [40]. While Eq. (2) is valid for arbitrary domain shapes, it requires domain walls free of internal singularities such as Bloch lines [33, 41]. Mathematically, $M(r)$ is a fiber bundle [42] on the base space $r$ and therefore locally flat but globally nontrivial [43]. In fact, Fig. 1(d-i) provides a simple example of a bulk-boundary equivalence, a feature that forms a cornerstone of topological physics [22]. The sign of $Q$ depends on the vorticity [44] of the spin structure, that is, on whether the region enclosed by the yellow boundary is red ($Q = +1$) or blue ($Q = -1$). In particular, $Q$ is independent of the clockwise or counterclockwise chirality of Bloch walls in the yellow region (Figs. S1, S2 in the Supplemental Material [45]).

As discussed above, TPTs do not increase or decrease order parameters but consist of changes in topological numbers. This leads to the question of whether such transitions lead to hysteretic features beyond magnetic hysteresis. This hysteresis was not recognized in earlier research, because available systems had micron-size rather than nanoscale feature sizes, which makes it very difficult to detect the Berry curvature in Hall-effect measurements. In this paper, we have fabricated exchange-coupled Co nanoparticle films having a much smaller average size of about 13.7 nm and show topological phase transitions and berry phase hysteresis using experiments. The experimental results and the underlying physics are also explained using micromagnetic simulations and an analytical model.

## II.    EXPERIMENTAL AND COMPUTATIONAL METHODS

An inert gas condensation-type cluster-deposition method, schematically shown in Fig. S3a in Supplement B, as described elsewhere [45, 46]. First, Co nanoparticles were produced by a DC magnetron sputtering using a mixture of argon and helium with a power of 200 W in a gas-aggregation chamber. After the formation, the nanoparticles were extracted towards the deposition chamber and deposited as a dense film on a Si (100) substrate having a Hall bar. The base pressure of the gas-aggregation chamber was $6 \times 10^{-8}$ Torr and the respective Ar and He flow rates were maintained at 400 and 100 SCCM (standard cubic centimeter per minute), respectively. The pressure in the cluster-formation chamber during the deposition was 0.7 Torr.

The Co nanoparticles were deposited with a low coverage density on a thin carbon film supported by copper grids for transmission-electron-microscopy measurements using an FEI Technai Osiris STEM. For magnetic and electron-transport measurements, the cluster-deposited nanoparticles were deposited for an extended time as a dense film as discussed in our previous works [46, 47]. The above measurements were performed using a superconducting quantum interference device (SQUID) and physical property measurement system (PPMS), respectively. A schematic of a dense nanoparticle film is shown in Fig S3(d) [45], and therefore they are exchange coupled and conducting. The thickness of the Co nanoparticle film is about 270 nm. The conduction channels for the Hall contacts were fabricated before depositing the Co nanoparticles, as described in Ref. 46. To prevent oxidation upon exposure to air, the Co nanoparticle film was capped with a $SiO_2$ layer of about 10 nm thickness immediately after deposition, using a radio-frequency magnetron sputtering. The $SiO_2$ cap layer is thinner (about 10 nm) as compared to the Co nanoparticle film of about 270 nm thickness and is also diamagnetic. Therefore, the film-$SiO_2$ interface is not expected to affect the magnetic and transport properties of the Co nanoparticle films. The particles have an average size of 13.7 nm with a narrow size distribution, see Fig S3 (b, c) and crystallize in the hcp structure, as shown in Fig. S4 [45]. A commercial AFM/MFM (Atto AFM/MFM Ixs; Attocube Systems) was used to map the topography and magnetic images at 200K. During the measurement, the MFM was performed in constant height mode (single pass) with PPP-MFMR tip from NANOSENSORS. The lift height is 250 nm and the scan speed is 5μm/s.

To numerically model the magnetic and Berry-phase hysteresis, we have performed micro-magnetic simulations using *ubermag* supported by OOMMF [48, 49]. We have numerically extracted skyrmion number $Q$ from the spin structure. A densely packed film of 1000 Co nanoparticles has been considered. The Co particles have sizes of about 13.7 nm and the total size of the simulated system, shown in Fig. S10, is 240 nm × 240 nm × 60 nm. We have used a computational cell size of 1.8 nm, which is well below the exchange length $l_{ex}$ [12], coherence radius 5.099 $l_{ex}$ of Co (10 nm) [12], and the domain-wall width (14 nm) of Co [12], and the current particle size. This cell size ensures a reasonable real-space resolution of $M(r)$.

Aside from the numerical cell size, our continuum approach is valid on length scales much larger than the Co-Co interatomic distance of 0.25 nm. This makes it possible to consider the thin film as a fiber bundle $M(r)$ with the base space $r$, allowing us to define quantities such as the boundary curvature $\kappa$. A refined atomistic analysis, not considered here, would yield corrections due to the discrete nature of the atoms at the particle's surfaces and near contact points, see e.g. Sect. 4.5 in Ref. 12. In particular, the crystal structure of Co is inversion-symmetric, so there are no bulk Dzyaloshinskii-Moriya interactions (DMI). Corrections to the magnetization angles caused by the DMI of surface atoms are likely but probably very small. Note that the Co nanoparticles exhibit nanoscale inversion symmetry, as contrasted, for example, to the magnetism of Co/Pt bilayers [27].

The cluster-deposition method yields isotropic nanoparticles with random grain orientation and therefore a random orientation of the easy magnetization axes $\bm{n}$ of the hcp Co particles, obeying $\langle n_x \rangle = \langle n_y \rangle = \langle n_z \rangle = 0$ and $\langle \bm{n}^2 \rangle = 1$. This randomness in simulations, clearly visible in Fig. S10(b), was implemented by using python np.random.uniform [45, 49]. The particles touch each other as shown in Fig S3(d), so that the exchange stiffness $A$ near the contact points is the same as in bulk Co.

Temperature-dependent micromagnetic effects are included in the lowest order, that is, by considering the intrinsic materials parameters $M_s$, $K_1$, and $A$ as temperature-dependent. This approach accounts for the atomic spin disorder outlined in Fig. 1(a). Other finite-temperature corrections, caused for example by magnetic viscosity [12], have been ignored. In our simulations, we have taken values of $M_s = 1300$ kA/m, $K_1 = 0.58$ MJ/m$^3$, and $A = 10.3$ pJ/m [12].

### III. RESULTS AND DISCUSSION

Transmission electron microscope and the corresponding particle-size histogram show an average particle size of 13.7 nm with a standard deviation $\sigma/d \approx 0.15$ (Figs. S3b and S3c) for the Co nanoparticles. We have conducted magnetic, electron-transport, and Hall-effect measurements at temperatures from 10 K to 300 K for the dense Co nanoparticle films as schematically shown in Fig. S 3(d). The magnetic hysteresis loops are shown in Fig. S5, and the measured coercivities are 0.18 T at 10 K and 0.04 T at 300 K.

Figure 2 compares the experimental data on a Co nanoparticle thin film (a) with numerical predictions (b). The THE was extracted from Hall-effect measurements, Fig. S6, as explained in Supplement B. We see that the Berry-phase hysteresis loops (colored) look qualitatively different from the magnetic hysteresis loops (black) and that they are much broader than the magnetic ones. Figure 2(a-b) also shows that Berry-phase hysteresis loops contain more features than magnetic hysteresis loops. There are both mathematical and physical explanations for these differences. Mathematically, Eq. (1) contains derivatives, which amounts to a numerical amplification of details. Physically, Berry-phase hysteresis loops exhibit a more complicated dependence on the spin structure, because $\Phi$ is more complicated than $\bm{m}$.

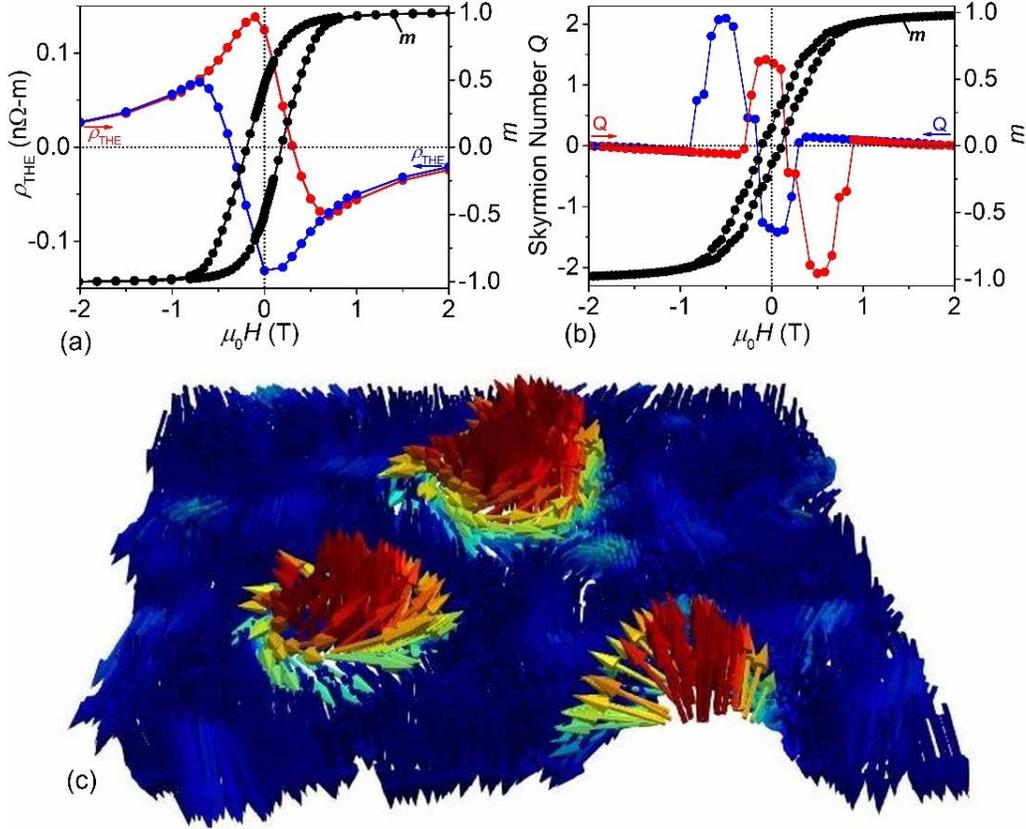

**Fig. 2. Magnetic hysteresis, Berry-phase hysteresis, and spin structure**: (a) experiment, (b) simulation, and (c) simulated spin structure in a field of -0.7 T. In (b), $Q$ is the number of skyrmions per unit area (240 nm × 240 nm × 60 nm), and $m$ is the normalized magnetization, $M_z/M_s$. The origin of the topological Hall effect due to spin texture, namely the noncoplanar spin structure, is visible in (c).

Magnetization reversal in a thin film of nanoparticle magnets is strongly real-structure dependent, which affects the magnetic [12, 50, 51] and, especially, Berry-phase hysteresis loops. There is an intricate balance between an interatomic exchange, magnetocrystalline anisotropy, and magnetostatic interactions, which result in spin structures such as that in Fig. 2(c) at some specific magnetic field. While the real structure leads to loop deformation, it does not affect the key feature of Figs. 1(d-i), namely red (or blue) regions in a blue (or red) background. This embedding is the origin of the THE. Note that the longitudinal resistivity of the Co nanoparticle thin film slightly increases with temperature, from 0.26 $\mu\Omega$ m at 10 K to 0.32 $\mu\Omega$ m at 300 K (Fig. S7). This shows that the film is metallic and that the Co nanoparticles touch each other. This metallic contact is necessary to ensure exchange coupling between the nanoparticles and a noncollinear spin structure like that in Fig. 2(c).

Figures 3(a-b) illustrate the mathematical effect, by comparing the $M(H)$ curve (black) with its field derivative, the micromagnetic susceptibility $\chi = dM/dH$ (purple). While the experimental and theoretical $M(H)$ curves [Fig. 3(a) and 3(b), respectively] look similar, the derivative greatly enhances the differences. In the present system, the susceptibility peaks are due to Barkhausen jumps [50, 52], which are strongly real-structure dependent. A schematic example of a Barkhausen jump is a fictitious transition from Fig. 1(d) to 1(f). These transition changes enhance the red area and therefore the magnetization in a jump-like fashion. In fact, there are two types of Barkhausen jumps, which have not yet been distinguished in the literature. When the field changes the domain

size and shape only, then $Q$ remains constant, but the Barkhausen-induced creation or merger of domains changes $Q$ like the transition from Fig. 1(f) to 1(g).

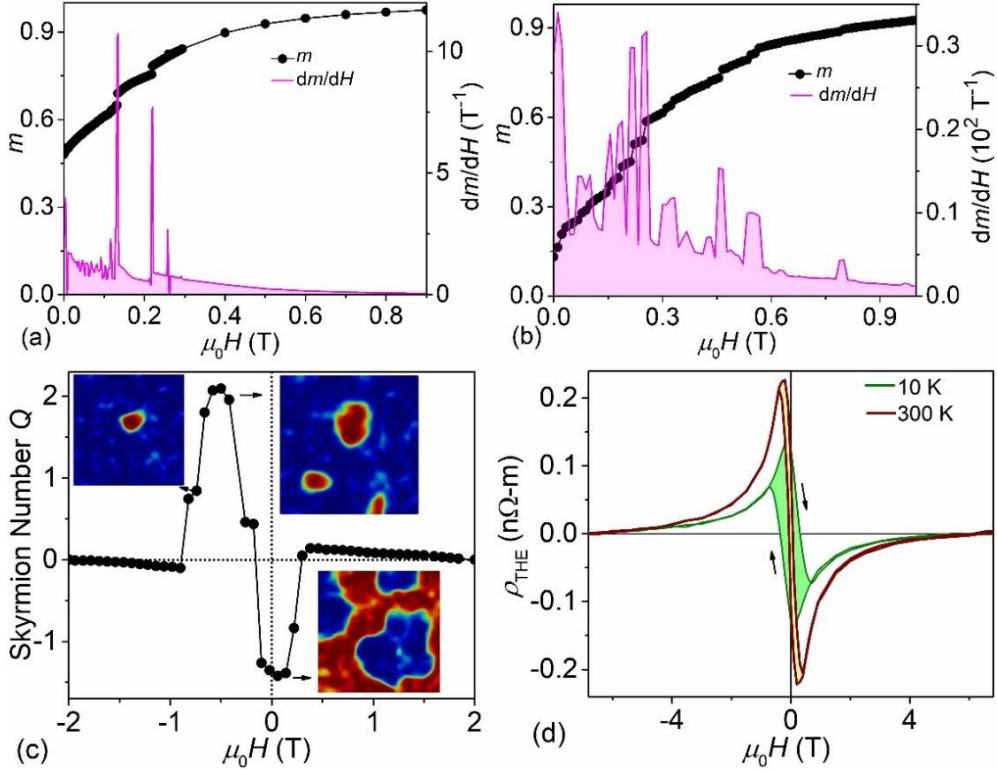

**Fig. 3. Real-structure and temperature effects**: (a) experimental magnetization and susceptibility, (b) simulated magnetization and susceptibility, (c) the real-structure origin of the Berry-phase hysteresis, and (d) Berry-phase hysteresis at 10 and 300 K. The susceptibility peaks in (a-b) reflect Barkhausen jumps and are strongly real-structure dependent.

Physically, the topological Hall signal critically depends on details of the magnetization process. Figure 3(c) outlines the situation in the present system, by showing how the red and blue areas evolve in a magnetic field. In strongly negative fields, the magnetization is ↓ (blue) everywhere, but with increasing field, the magnetization starts to become noncoplanar noncollinear, and Eq. (1) yields a nonzero skyrmion density. The red regions grow and finally coalesce. This coalescence does not change the magnetization very much but yields a drastic change in $Q$: red regions in a blue background become blue regions in a red background, which causes the sign of $Q$ to switch. Since we investigate inhomogeneous nanoparticle thin film, the switching patterns exhibit considerable randomness, Fig. 3(c), but this does not affect the overall topological picture.

The magnetization and magnetization reversal in the exchange-coupled Co nanoparticle film can be explained using the magnetization reversal process, as schematically shown in Fig. S15. A key question is whether the reversal is noncooperative as in Figure S15(a), as compared to a cooperative reversal in Figure S15(b). It is known that noncooperative reversal dominates in systems with broad switching-field distributions (small values of *dM/dH*), whereas narrow switching field distributions (large values of d*M*/d*H*) favor cooperative reversal [12]. The underlying physical mechanism is that the interatomic exchange proportional to $A/R^2$, where $A$ is the exchange stiffness, competes against the anisotropy of strength $K_1$. For very small particles, as well as in very soft magnets (small $K_1$), the exchange dominates and the reversal is cooperative. Elemental Co is a prototypical semihard magnet and particle size is fairly small, so we are in an intermediate regime closer to cooperative reversal, Fig. S15(b) than to noncooperative reversal, Fig.S15(a). The sizes of the cooperative blocks are random, and some of the blocks are fairly large. The switching of such big blocks has the character of Barkhausen jumps.

Another intriguing aspect is an increase in the topological Hall effect with increasing temperature. As mentioned above, the Co nanoparticles are exchanged coupled, which exhibit cooperative magnetization reversal and subsequently gives rise to the chiral domains with chiral domain walls i.e yellow boundary enclosing the region with uniform magnetization (Figure S1). As temperature increases, the exchange stiffness constant decreases in Co [See supplemental material for calculation of exchange stiffness in Co nanoparticles and Refs. 53, 54, 55 for more details]. Second, at elevated temperatures, the reversal is not only accompanied by nucleation but also by the thermal fluctuation of spins i.e., the probability of flipping of spins increases as temperature increases because of a small increase in thermal energy (Ref. 55, Ch. 6). These two factors cause the magnetization reversal easier and lead to an increase in the number of individual domains with the noncoplanar spin structures. Therefore, an increase in temperature is expected to increase the intensity of the topological Hall effect as observed in the case of exchange-coupled Co nanoparticle film. This feature, linked to the high Curie temperature of Co, is an advantage because the non-coplanar spin texture with finite skyrmion number caused by B20 Dzyaloshinskii-Moriya interactions requires considerable effort at high temperatures [38]. It is worth noting that the chiral domains reported in this study are only quantified in terms of skyrmion numbers and are not traditional skyrmions caused by Dzyaloshinskii–Moriya interaction (DMI) in B20-type materials and interfacial DMI in multi-layered thin films.

It is worth noting that anomalies in Hall-effect may also arise from two or multichannel Hall effects [56, 58, 59]. However, the two-channel Hall effect discussed in [56, 58, 59] can be ruled out in the exchange-coupled Co nanoparticle film. First, the two-channel Hall effect in the heterostructures discussed in Ref. 58 is solely due to the presence of the heavy metal Pt layer, which causes the spin Hall effect, and also AHE due to proximity-induced magnetization. These effects are not expected in homogeneous Co nanoparticle films. Second, inhomogeneous systems with the presence of a secondary magnetic phase often show bimodal switching field distributions in magnetization data (For example, Fig. 4(d) in Ref. 56 and 57) and subsequently yield the two-channel hall effect. This bimodal two-channel effect generally arises in systems with combinations of hard and soft phases, which will not give narrow switching-field distribution as discussed in Ref. 12. In contrast, our Co nanoparticle film does not show any bimodal switching field distributions in the magnetization data Fig. 2(a) in the main manuscript and Fig. S5 in the supplemental material). Finally, in the case of Mn-doped $Bi_2Se_3$ discussed in Ref. 59, the two-channel Hall effect is caused by different magnetization contributions from bulk and surface, which is not the case in the exchange-coupled Co nanoparticle films. With all physical origins that cause the two-channel Hall effect [56, 58, 59] ruled out in the exchange-coupled Co nanoparticle films, the anomalies observed in the Hall data are solely due to the topological Hall effect.

We also measured the change of magnetic domains of the Co nanoparticle film by magnetic force microscope (MFM) (see Figure 4) and the corresponding Atomic Force Microscopy (AFM) topography images at room temperature [45]. As discussed above, the exchange-coupled nanoparticles involve in cooperative magnetization reversal, and the reversed magnetic domains expand with increasing the magnetic field. This is seen from the phase images of MFM, which show that the individual magnetic domains with closed domain walls appear at around 0.04 T, and their size increases as the magnetic field increases. Our electron-transport data show that the THE has a maximum value in the region -0.02 T to 0.02 T. The MFM in images also shows a comparatively large number of smaller magnetic bubbles around this field region. Note that the bigger magnetic domains may still contain several small domains in the field region -0.02 T to 0.02 T, which could not be visualized due to the low resolution of MFM.

In the MFM images, the positive (negative) phase shift corresponds to the repulsive (attractive) force between the tip and magnetic stray field. When the sample is fully magnetized at a saturated field, the parallel alignment of the magnetic moment for the sample and tip should contribute to the negative phase. However, under lift mode, the severe change of surface roughness may perturb the phase signal, and there exist areas with a positive phase (the yellow region in the uneven surface) that persist even under 5 T. However, in a relatively flat area, the closed magnetic domains with negative phase signals undergo field-driven expanding and subsequent coalescence. Our AFM and MFM images are shown in Figs. 4, S13 and S14 show that the region in which the field gradient disappear is almost even, while the region where the field gradient does not change has an uneven surface.

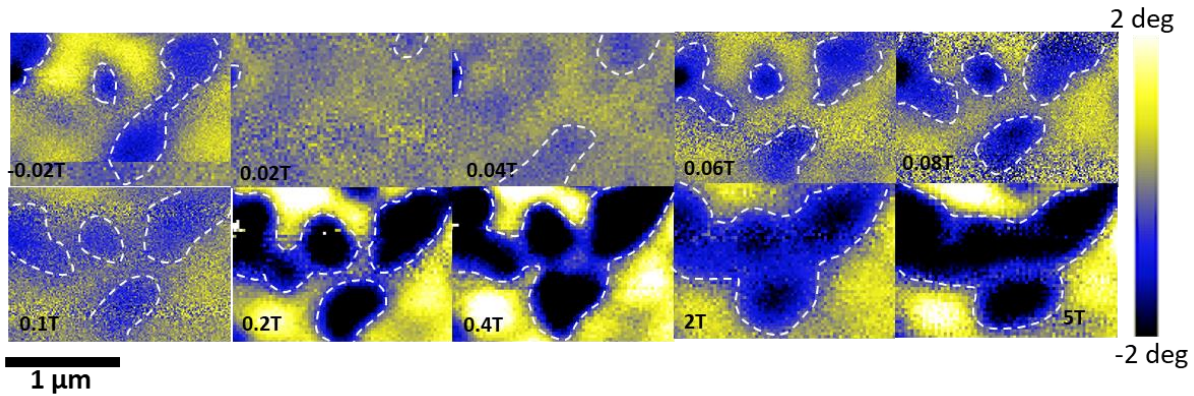

**Fig. 4. MFM image and magnetization reversal**: MFM image of magnetization reversal in a region with an even surface (blue region). The magnetization reversal starts randomly making domains at a field of 0.06 T and these domains expand as the field is increased and finally coalescence of the domain occurs at a high field. Note that not only do we have topological contributions due to magnetic domains, but we also have a topological contribution to THE due to chiral spin inhomogeneity, and imaging of those spins is difficult [47, 60].

Magnetic hysteresis loops are typically plotted by showing the magnetization $M$ as a function of the magnetic field $H$. Figures 2 and 3 show that this is also possible for Berry-phase hysteresis. However, such plots convolute magnetic and topological properties, because the field generally changes both $M$ and its gradient $\nabla M$. To remove this field effect, we introduce a new plot showing $Q$ as a function of $m = M/M_s$ (Fig. 5). In this parametric $Q$-$M$ plot, each field corresponds to one point in $Q$-$M$ space, but this field is not shown explicitly unless each point of the curve is explicitly labeled by its field value.

The $Q$-$M$ plot provides not only an entirely new view of Berry-phase hysteresis but also simplifies the analysis. In particular, $m = -1$ and $m = +1$ correspond to homogeneous magnetization states, so that $Q(m = \pm 1) = 0$. Berry-phase hysteresis occurs for intermediate values of $m$, and Fig.

5(a) shows that this hysteresis is accompanied by a topological remanence $Q_o$. Approximating the $Q(m)$ by a cubic polynomial [61] yields

$$Q = (Q_o + Q_1 m)(1 - m^2) \qquad (3)$$

A cubic polynomial contains four parameters, but only two of them, namely $Q_o$ and the magnitude-parameter $Q_1$, are adjustable. The remaining two parameters are implicitly fixed by the boundary conditions at $m = \pm 1$. While Eq. (3) is a rather crude approximation, it works surprisingly well for the present system, as evidenced by the comparison with experiment in Fig. 5(b).

Figures 5(c-g) show how the Berry-phase hysteresis evolves in a simple exactly solvable model. Circular red domains are arranged on a triangular lattice (c) and grow in the external field (c). At the phase-transition point, the domains touch and start to overlap, so that the background changes from blue to red and the THE changes sign. Figure 5(g) shows the $Q$-$m$ plot for the model of (c-d). The transition is triggered by an external magnetic field, but near the transition point, there is only a trivially small magnetization change. We also note that the transition (c→d) occurs at a point where most of the area is red (↑) already, at a magnetization of $\pi/\sqrt{3} - 1 \approx 0.814\, M_s$. This is the reason for the striking width of the green topological hysteresis loop in Fig. 5(g).

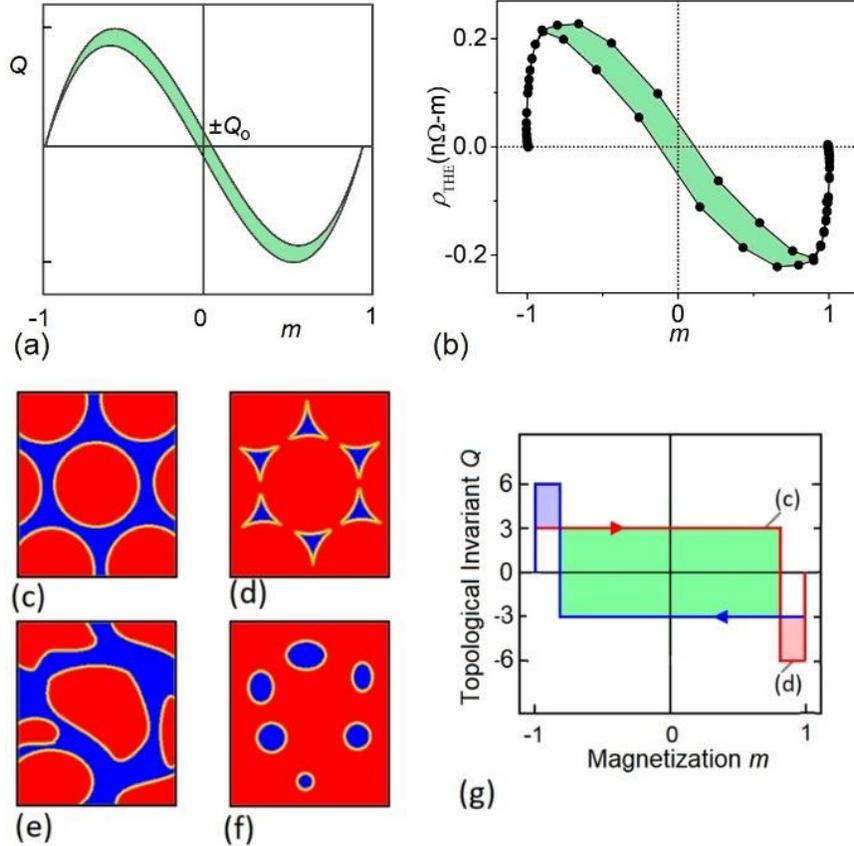

**Fig. 5.** *Analytical modeling of Berry-phase hysteresis*: (a) most general cubic plot of skyrmion number $Q$ as a function of magnetization, (b) experimental $Q$-$M$ plot, (c-d) topological phase transition (TPT) in a simple circular-domain model, (e-f) topologically equivalent version of the same model, and (g) Berry-phase hysteresis loop for the model of (c-d). The transitions from (c) to (d) and from (e) to (f) are triggered by a magnetic field increase and accompanied by an incremental magnetization increase only. In (g), the transition occurs at a fairly high value of $M_z = \pm 0.814\, M_s$. At $M_z = \pm M_s$, $Q$ jumps to zero, because the residual domains are annihilated at saturation. The light-blue and light-red areas in (g) are a duality effect caused by the triangular skyrmion lattice assumed in (a-b). As in other parts of the paper, $m$ is the normalized magnetization ($m = M_z/M_s$).

The light blue and light red areas in Fig. 5(g) are model-specific and related to the duality of the assumed skyrmion lattice. Figures 5(c) and 5(d) correspond to triangular and honeycomb lattices, respectively, which are dual but have different numbers of sites per honeycomb unit cell (3 and 6). By comparison, square lattices are self-dual, which causes the bright areas to disappear. In the light of this model analysis, the difference between the red/blue curves in Figs. 2(a) and (b) is not surprising, but a comprehensive explanation of the duality effect is a challenge to future mathematical and physical research. The domain structures with domain-wall chirality of Figs. 5(e-f) are topologically but not micromagnetically equivalent to (c-d) and yield a real-structure dependent smoothing of the rectangular loop parts in (g).

It is interesting to note that domain structures like those in Figs. 1 and 5 have been around for decades [50, 62], as it has the recognition of features such as domain-wall chirality [62, 63]. However, at that time, neither the Berry phase nor the topological Hall effect was widely aware [5, 32, 37]. Furthermore, the initial research focused on bubble domains of fairly large sizes $L$, typically micron-sized. Each reverse domain contributes one flux quantum to the THE, so that the net effect scales a $1/L^2$. Even today, such small effects are nontrivial to detect without the help of Dzyaloshinski-Moriya (DM) interactions [17], and this is the main reason for our consideration of Co nanoparticle thin films, where $L$ a few 10 nm.

Future applications of Berry-phase hysteresis in spin electronics and beyond [64, 65, 66] are difficult to judge. The low skyrmion mobility in the present nanoparticle system will probably prevent applications such as racetrack memories [64], but three arguments speak in favor of the potential technological usefulness of Berry-phase hysteresis. First, the small feature size addresses miniaturization requirements in spin electronics. Second, Co has a very high Curie temperature, which facilitates the measurement and practical exploitation of its THE. In fact, Fig. 3(d) shows that the effect actually *increases* with temperature. By contrast, noncentrosymmetric materials tend to have rather low magnetic ordering temperatures, requiring considerable effort to drive the systems beyond room temperature [23, 38]. Third, the effect has a very high field sensitivity, as one can see from Fig. 3(d), and by comparing the maximum slopes of the red, blue, and black curves in Fig. 2(b).

## IV. CONCLUSIONS

The starting point and first main finding in this paper is the recognition that the thin-film magnetization reversal has the character of a topological phase transition. The transition is accompanied by Berry-phase hysteresis, a phenomenon very different from ordinary magnetic hysteresis and exhibiting features such as topological remanence and micromagnetic duality. The new concept has led to the development of a topology-specific plot showing the skyrmion number as a function of magnetization rather than the field. In our Co nanoparticle system, the Berry-phase hysteresis is realized on a nanoscale, increases with temperature, and exhibits a high field sensitivity. Several interdisciplinary challenges emerge from the present work. For example, it is intriguing to see which other systems investigated in the past, presence, and future exhibit Berry phase hysteresis and how it is realized.

**Acknowledgments**

This research is primarily supported by NSF-EQUATE under Grant No. OIA-2044049 (micromagnetic simulations) and the NU Collaborative Initiative (fabrication and characterization). This work was performed in part in the Nebraska Nanoscale Facility and Nebraska Center for Materials and Nanoscience, which are supported by the National Science

See Supplemental Material for additional details on methods and results [45], which also include Refs. [67, 68, 69, 70].

**References**

[1] J. M. Kosterlitz, Rev. Mod. Phys. **89**, 040501-1-7 (2017).

[2] Ch. Lin, M. Ochi, R. Noguchi, K. Kuroda, M. Sakoda, A. Nomura, M. Tsubota, P. Zhang, C. Bareille, K. Kurokawa, Y. Arai, K. Kawaguchi, H. Tanaka, K. Yaji, A. Harasawa, M. Hashimoto, D. Lu , Sh. Shin, R. Arita, S. Tanda, and T. Kondo, Nat. Mater. **20**, 1093-1099 (2021).

[3] M. Levin and X.-G. Wen, Phys. Rev. Lett. **96**, 110405-1-4 (2006).

[4] X.-G. Wen, *Quantum-Field Theory of Many-Body Systems*, University Press, Oxford 2004.

[5] D. Xiao, M.-Ch. Chang, and Q. Niu, Rev. Mod. Phys. **82**, 1959–2007 (2010).

[6] B. A. Bernevig, T. L. Hughes, and S.-C. Zhang, Science **314**, 1757-1761 (2006).

[7] I. M. Lifshitz, Sov. Phys. JETP **11,** 1130-1135 (1960).

[8] J. M. Kosterlitz and D. J. Thouless, J. Phys. C **6**, 1181-1203 (1973).

[9] H. Huang and F. Liu, AAAS Research **2020**, 7832610-1-7 (2020).

[10] E. Warburg, Ann. Phys. (Leipzig) **249**, 141-164 (1881).

[11] M. Dörries, Hist. Stud. Nat. Sci. **22**, 25–55 (1991).

[12] R. Skomski, J. Phys.: Condens. Matter. **15**, R841-896 (2003).

[13] L. D. Barron, "An introduction to Chirality at the Nanoscale", in: *Chirality at the Nanoscale*, Ed. D. B. Amabilino, Wiley, New York 2009, p. 1-27.

[14] W. Jiang, P. Upadhyaya, W. Zhang, G. Yu, M. B. Jungfleisch, F. Y. Fradin, J. E. Pearson, Y. Tserkovnyak, K. L. Wang, O. Heinonen, S. G. E. te Velthuis, and A. Hoffmann, Science **349**, p. 283-286 (2015).

[15] M. Lonsky and A. Hoffmann, Phys. Rev. B **102**, 104403-1-11 (2020).

[16] S.-G. Je, H.-S. Han, S. K. Kim, S. A. Montoya, W. Chao, I.-S. Hong, E. E. Fullerton, K.-S. Lee, K.-J. Lee, M.-Y. Im, and J.-I. Hong, ACS Nano **14**, 3251-3258 (2020).

[17] W. Zhang, B. Balamurugan, A. Ullah, R. Pahari, X. Li, L. Yue, S. R. Valloppilly, A. Sokolov, R. Skomski, and D. J. Sellmyer, Appl. Phys. Lett. **115**, 172404-1-4 (2019).

[18] W. Wang, Y.-F. Zhao, F. Wang, M. W. Daniels, C.-Z. Chang, J. Zang, D. Xiao, and W. Wu, Nano Lett. **21**, 1108-1114 (2021).

[19] P. Li, J. Ding, S. S.-L. Zhang, J. Kally, T. Pillsbury, O. G. Heinonen, G. Rimal, Ch. Bi, A. DeMann, S. B. Field, W. Wang, J. Tang, J. S. Jiang, A. Hoffmann, N. Samarth, and M. Wu, Nano Lett. **21**, 84-90 (2021).

[20] E. Y. Vedmedenko, R. K. Kawakami, D. D. Sheka, P. Gambardella, A. Kirilyuk, A. Hirohata, C. Binek, O. Chubykalo-Fesenko, S. Sanvito, J. Grollier, K. Everschor-Sitte, T Kampfrath, C.-Y. You, and A Berger, J. Phys. D: Appl. Phys. **53**, 453001-1-44 (2020).

[21] Y. Taguchi, Y. Oohara, H. Yoshizawa, N. Nagaosa, and Y. Tokura, Science **291,** 2573-2576 (2001).

[22] M. Z. Hasan and C. L. Kane, Rev. Mod. Phys. **82**, 3045-3067 (2010).

[23] Y. Tokunaga, X. Z. Yu, J. S.White, H. M. Rønnow, D. Morikawa, Y. Taguchi, and Y. Tokura, Nat. Commun. **6**. 7638-1-13 (2015).

[24] K. G. Rana, O. Meshcheriakova, J. Kübler, B. Ernst, J. Karel, R Hillebrand, E. Pippel, P. Werner, A. K. Nayak, C. Felser, and S. S. P. Parkin, New J. Phys. **18**, 085007 (2016).

[25] A. Soumyanarayanan, N. Reyren, A. Fert, and Ch. Panagopoulos, Nature **539**, 509-517 (2016).

[26] D. S. Sanchez, I. Belopolski, T. A. Cochran, X. Xu, J.-X. Yin, G. Chang, W. Xie, K. Manna, V. Süß, Ch.-Y. Huang, N. Alidoust, D. Multer, S. S. Zhang, N. Shumiya, X. Wang, G.-Q. Wang, T.-R. Chang, C. Felser, S.-Y. Xu, Sh. Jia, H. Lin, and M. Z. Hasan, Nature **567**, 500-499 (2019).

[27] M. Raju, A. Yagil, A. Soumyanarayanan, A. K. C. Tan, A. Almoalem, Fusheng Ma, O. M. Auslaender, and C. Panagopoulos, Nat. Commun. **10**, 696-1-7 (2019).

[28] A. K. Sharma, J. Jena, K. G. Rana, A. Markou, H. L. Meyerheim, K. Mohseni, A. K. Srivastava, I. Kostanoskiy, C. Felser, and S. S. P. Parkin, Adv. Mater., 33, 2101323 (2021).

[29] M. Tomé and H. D. Rosales, Phys. Rev. B **103**, L020403-1-5 (2021).

[30] J. P. Pekola, K. Torizuka, A. J. Manninen, J. M. Kyynäräinen, and G. E. Volovik, Phys. Rev. Lett. **65**, 3293-3296 (1990).

[31] N. Nagaosa and Y. Tokura, Nat. Nanotechnol. **8**, 899-911 (2013).

[32] A. Neubauer, C. Pfleiderer, B. Binz, A. Rosch, R. Ritz, P. G. Niklowitz, and P. Böni, Phys. Rev. Lett. **102**, 186602-1-4 (2009).

[33] S. Seki and M. Mochizuki, "Skyrmions in magnetic materials", Springer International, Cham (2016).

[34] Y. Fujishiro, N. Kanazawa, T. Nakajima, X.Z. Yu, K. Ohishi, Y. Kawamura, K. Kakurai, T. Arima, H. Mitamura, A. Miyake, K. Akiba, M. Tokunaga, A. Matsuo, K. Kindo, T. Koretsune, R. Arita, and Y. Tokura, Nat. Commun. **10**, 1059 1-8 (2019).

[35] L. Pierobon, Ch. Moutafis, Y. Li, and J. F. Löffler, Sci. Rep. **8**, 16675-1-9 (2018).

[36] S. Das, Z. Hong, V. A. Stoica, M. A. P. Gonçalves, Y. T. Shao, E. Parsonnet, E. J. Marksz, S. Saremi, M. R. McCarter, A. Reynoso, C. J. Long, A. M. Hagerstrom, D. Meyers, V. Ravi, B. Prasad, H. Zhou, Z. Zhang, H. Wen, F. Gómez-Ortiz, P. García-Fernández, J. Bokor, J. Íñiguez, J. W. Freeland, N. D. Orloff, J. Junquera, L. Q. Chen, S. Salahuddin, D. A. Muller, L. W. Martin, and R. Ramesh, Nat. Mater. **20**, 194-201 (2021).

[37] M. V. Berry, Proc. R. Soc. Lond. A **392**, 45-57 (1984).

[38] B. Balasubramanian, P. Manchanda, R. Pahari, Z. Chen, W. Zhang, S. R. Valloppilly, X. Li, A. Sarella, L. Yue, A. Ullah, P. Dev, D. A. Muller, R. Skomski, G. C. Hadjipanayis, and D.J. Sellmyer, Phys. Rev. Lett. **124**, 057201-1-6 (2020).

[39] R. Skomski, B. Balasubramanian, A. Ullah, C. Binek, and D. J. Sellmyer, AIP Adv., **12**, 035341 (2022).

[40] W. Fenchel, "Über Krümmung and Windung geschlossner Raumkurven", Math. Ann. **101**, 238-252 (1929).

[41] Y. He, T. Helm, I. Soldatov, S. Schneider, D. Pohl, A. K. Srivastava, A. K. Sharma, J. Kroder, W. Schnelle, R. Schaefer, B. Rellinghaus, G. H. Fecher, S. S. P. Parkin, and C. Felser, Phys. Rev. B **105**, 064426-1-8 (2022)

[42] H. Seifert, "Topologie dreidimensionaler gefaserter Räume", Acta Math. **60**, 147–238 (1933).

[43] J. W. Zwanziger, M. Koenig, and A. Pines, "Berry's Phase", Annu. Rev. Phys. Chern. **41**, 601-646 (1990).

[44] B. Ding, J. Zhang, H. Li, S. Zhang, E. Liu, G. Wu, X. Zhang, and W. Wang, Appl. Phys. Lett. **116**, 132402-1-5 (2020).

[45] See Supplemental Material at http://link.aps.org/supplemental/ for additional details on methods and additional results.

[46] B. Balasubramanian, T. A. George, P. Manchanda, R. Pahari, A. Ullah, R. Skomski, and D. J. Sellmyer, Phys. Rev. Mater. **5**, 024402-1-9 (2021).

[47] R. Pahari, B. Balasubramanian, A. Ullah, P. Manchanda, H. Komuro, R. Streubel, C. Klewe, Shah R. Valloppilly, P. Shafer, P. Dev, R. Skomski, and David J. Sellmyer, Phys. Rev. Mater. **5**, 124418 (2021).

[48] M. Beg, R. A. Pepper, and H. Fangohr. AIP Advances **7**, 56025 (2017), M. Beg, R. A. Pepper, T. Kluyver, J. Mulkers, J. Leliaert, and H. Fangohr. ubermag: Meta package for Ubermag project. DOI: 10.5281/zenodo.3539496 (2021).

[49] Our code is publically available: https://github.com/ubermag/help/issues/110#issuecomment-970591821

[50] A. Hubert and R. Schäfer, *Magnetic Domains*, Springer-Verlag, Berlin (1998).

[51] L. Exl, D. Suess, and T. Schrefl, "Micromagnetism", in: *Handbook of Magnetism and Magnetic Materials*, Eds. J. M. D. Coey and S. S. Parkin, Springer, Cham, p. 347-390.

[52] H. Barkhausen, "Zwei mit Hilfe der Neuen Verstärker entdeckte Erscheinungen", Phys. Z. **20**, 401-403 (1919).

[53] H. Kronmüller and M. Fähnle, Micromagnetism and the Microstructure of Ferromagnetic Solids (University Press, Cambridge 2003).

[54] R. Moreno, R. F. L. Evans, S. Khmelevskyi, M. C. Munoz, R. W. Chantrell, and O. Chubykalo-Fesenko, Phys. Rev. B **94**, 104433 (2016).

[55] J. M. D. Coey, Magnetism and Magnetic Materials (Cambridge University Press, New York, 2010), R. Skomski and J. M. D. Coey, Permanent Magnetism (Institute of Physics, Bristol, 1999).

[56] G. Kimbell, P. M. Sass, B. Woltjes, E. K. Ko, T. W. Noh, W. Wu, and J. W. A. Robinson, Phys. Rev. Mater. **4**, 054414 (2020).

[57] A. Gerber, Phys. Rev. B **98**, 214440 (2018).

[58] S. Ding, Z. Liang, C. Yun, R. Wu, M. Xue, Z. Lin, A. Ross, S. Becker, W. Yang, X. Ma, D, Chen, K. Sun, G. Jakob, M. Kläui, and J. Yang, Phys. Rev. B **104**, 224410 (2021).

[59] N. Liu, J. Teng, and Y. Li, Nat. Commun. **9**, 1282 (2018).

[60] Haohan Wang, Balamurugan Balasubramanian, Rabindra Pahari, Ralph Skomski, Yaohua Liu, Ashfia Huq, D. J. Sellmyer, and Xiaoshan Xu, Phys. Rev. Mater. **3**, 064403 (2019).

[61] W. Zhang, B, Balasubramanian, Y. Sun, A. Ullah, R. Skomski, R. Pahari, S. R. Valloppilly, X.-Zh. Li, C.-Zh. Wang, K.-M. Ho, and D. J. Sellmyer, J. Magn. Magn. Mater. **537**, 168104 (2021).

[62] P. J. Grundy and S. R. Herd phys. stat. sol. (a) **20**, 295-307 (1973).

[63] J. Jiang, D. Xiao, F. Wang, J.-H. Shin, D. Andreoli, J. Zhang, R. Xiao, Y.-F. Zhao, M. Kayyalha, L. Zhang, K. Wang, J. Zang, C. Liu, N. Samarth, M. H. W. Chan, and C.-Z. Chang, Nat. Mater. **19**, 732 (2020).

[64] S. S. P. Parkin, M. Hayashi, and L. Thomas, Science **320**, 190-194 (2008).

[65] A. Fert, N. Reyren, and V. Cros, Nat. Rev. Mater. **2**, 17031-1-15 (2017).

[66] C. Back, V. Cros, H. Ebert, K. Everschor-Sitte, A. Fert, M. Garst, T. Ma, S. Mankovsky, T. L. Monchesky, M. Mostovoy, N. Nagaosa, and S. S. P. Parkin, J. Phys. D **53**, 363001-1-37 (2020).

[67] A. A. Thiele, Bell Syst. Tech. J. **48**, 3287 (1969).

[68] W. Koshibae, N. Nagaosa, New J. Phys. **18** 045007 (2016).

[69] B. Balasubramanian, X. Zhao, S. R. Valloppilly, S. Beniwal, R. Skomski, A. Sarella, Y. Jin, X. Li, X. Xu, H. Cao, H. Wang, A. Enders, C.-Z. Wang, K.-M. Ho, and D.J. Sellmyer, Nanoscale, **10**, 13011 (2018).

[70] X. H. Zhang, T. R. Gao, L. Fang, S. Fackler, J. A. Borchers, B. J. Kirby, B. B. Maranville, S. E. Lofland, A. T. N'Diaye, E. Arenholz, A. Ullah, J. Cui, R. Skomski, I. Takeuchi J. Magn. Magn. Mater. **560**, 169627 (2022).

# Supplemental Materials
# Topological Phase Transitions and Berry-Phase Hysteresis in Exchange-Coupled Nanomagnets

Ahsan Ullah,[*] Xin Li, Yunlong Jin, Rabindra Pahari, Lanping Yue, Xiaoshan Xu, Balamurugan Balasubramanian, David J. Sellmyer,[‡] and Ralph Skomski[‡]

*Department of Physics & Astronomy and Nebraska Center for Materials and Nanoscience, University of Nebraska, Lincoln, NE 68588*

*E-mail: aullah@huskers.unl.edu
[‡]Deceased

### Supplement A: Theoretical evaluation of the skyrmion density

The topological Hall effect is measured in the film plane (*x*-*y*-plane), so that its prediction for a given micromagnetic spin structure $\mathbf{M}(\mathbf{r})$ requires the integration $Q = \int \Phi \, dx \, dy$, where $\Phi$ depends on $\partial \mathbf{m}/\partial x$ and $\partial \mathbf{m}/\partial y$, Eq. (1). There are also nonzero derivatives $\partial \mathbf{m}(\mathbf{r})/\partial z$, but these do not contribute to the THE due to the measurement geometry. Inside the red and blue regions, the magnetization is therefore constant, so that $\Phi = 0$. We can therefore restrict the evaluation of Figs. 1(d-i) and 4(c-f) to the yellow boundary region.

It is convenient to divide the integration into two parts, namely perpendicular and parallel to the yellow boundary. The mathematical justification for this approach is that fiber bundles $M(r)$ forms a locally flat space, irrespective of the global topological situation [1]. The integration perpendicular to the boundary yields a vorticity [2, 3] factor that is proportional to $\int \sin\theta \, d\theta$ but independent of the spin direction (chirality) explained in Fig. S1. Note that there is a key distinction between vorticity and chirality. Vorticity corresponds to the ↑ or ↓ magnetization in the core, whereas chirality refers to the left- or right-handiness of spins around the core. Since the red and blue regions have $m = \pm 1$, the vortex integration produces a factor of 2, and the subsequent integration along the boundary yields Eq. (2).

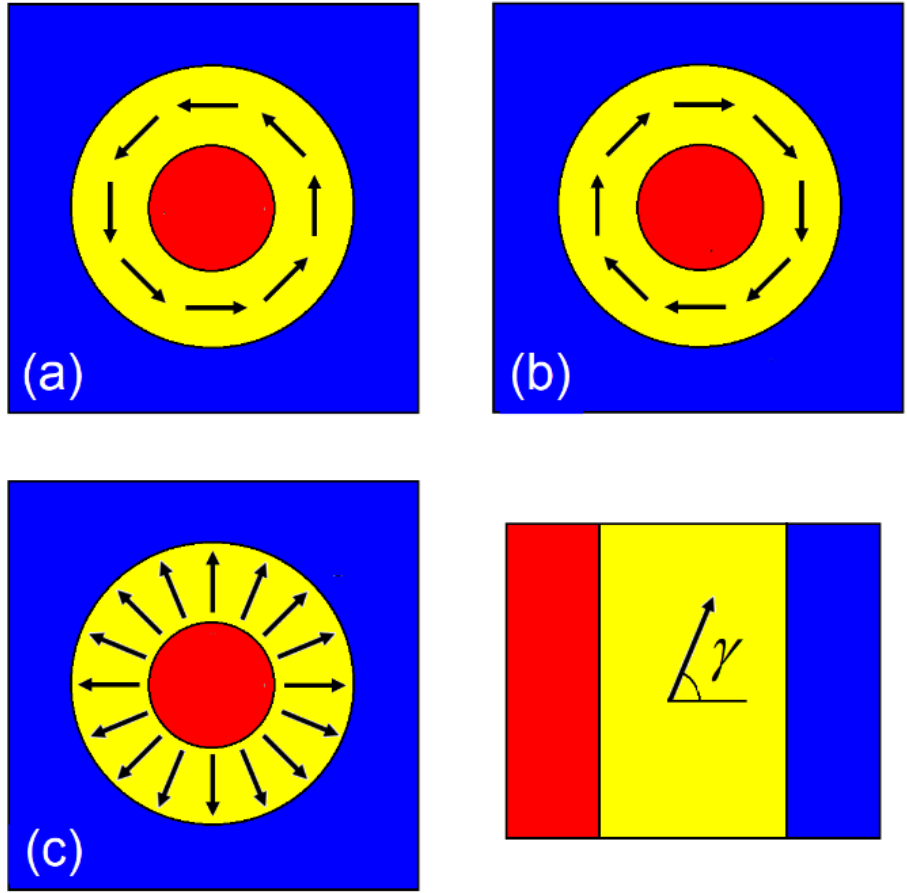

**Fig. S1. Chirality and vorticity**: (a) Bloch wall with counterclockwise chirality, (b) Bloch wall with clockwise chirality, (c) Néel wall, and (d) angle $\gamma$ of the spin direction. The THE depends on the vorticity only (red on blue background or blue on red background) but not on the angle $\gamma$, so that (a-c) yield $Q = +1$.

Circular geometries have $\kappa = 1/R$, so the integration in $Q = {}^1\!/_{2\pi} \oint \kappa \cdot d\mathbf{l}$ is trivial and yields $Q = \pm 1$. Figure S2 outlines the situation for arbitrary domain walls free of singularities such as Bloch lines. The sign of the local curvature depends on whether the focal point lies in the red area ($R$) or in the blue area ($R'$). Locally, it is not possible to determine the sign of $Q$, but globally it is, because $\oint \boldsymbol{\kappa} \cdot \mathrm{d}\boldsymbol{l} = 2\pi$ [4]. This integration elucidates that $Q$ depends on vorticity [2, 3] only, that is, on whether a red area (↑) is surrounded by a blue area (↓) or vice versa. The topological number does not change as a domain grows or gets distorted.

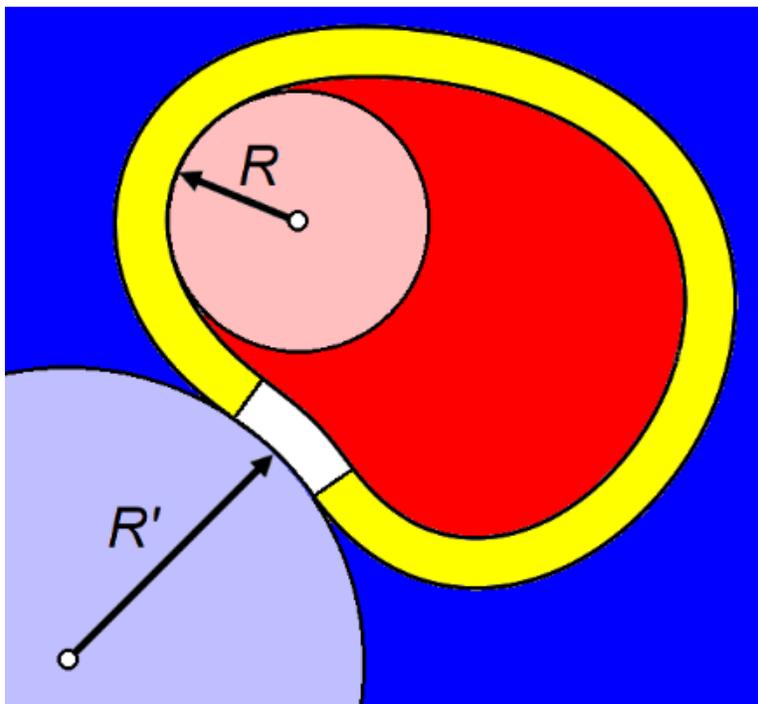

**Fig. S2.** Domain wall curvature and topological Hall effect. Depending on the location of the focal point, the local THE contribution may be positive (yellow part of the boundary) or negative (white part of the boundary). However, the contour integral along the entire boundary is topologically protected and equal to $2\pi$ in the present figure.

## Supplement B. Experimental and computational methods and analysis

Figure S3 explains the inert-gas condensation-type cluster-deposition machine used to fabricate and deposit hcp Co nanoparticles. The base pressure of the gas-aggregation chamber is $6 \times 10^{-8}$ Torr and the flow rates of Ar and He were maintained at 400 and 100 SCCM (standard cubic centimeter per minute), respectively. The pressure in the cluster-formation chamber during the deposition was 0.7 Torr. The transmission electron microscope results on Co nanoparticles are shown in Figs. S3(b-c). Note that the cluster-deposition method is a well-established method in producing nanoparticles in the gas phase (See supporting information of Balasubramanian *et al*. Advanced Materials, **25**, 6090 (2013) and ACS Nano **4**, 1893 (2010) for examples of cross-sectional images of cluster-deposited nanoparticles).

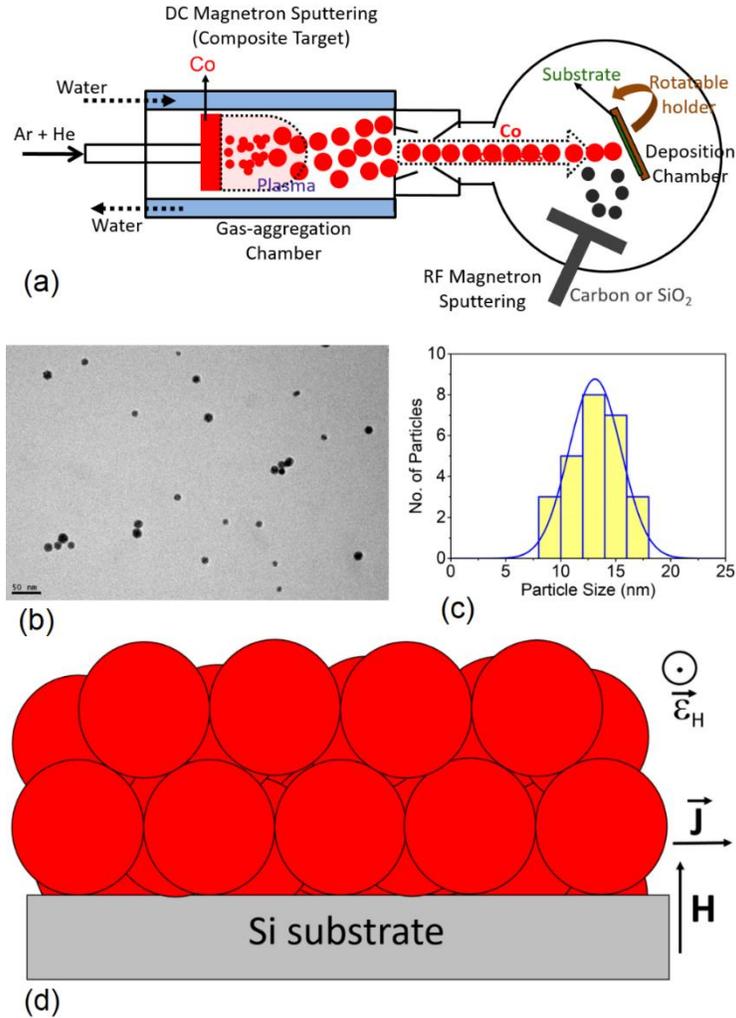

**Fig. S3** (a) A schematic of the cluster-deposition system. (b) Transmission electron microscope image and (c) the corresponding particle-size histogram. The result shows an average particle size $d = 13.7$ nm and a standard deviation $\sigma/d \approx 0.15$. (d) A schematic of a dense Co nanoparticle film used for magnetic and transport measurements.

Figure S4 shows the x-ray diffraction (XRD) pattern, which confirms that the nanoparticles are predominantly hcp Co by showing the intense (hkl) peaks corresponding to hcp structure (black). A few very weak peaks corresponding to cubic Co (green) is also observed [69].

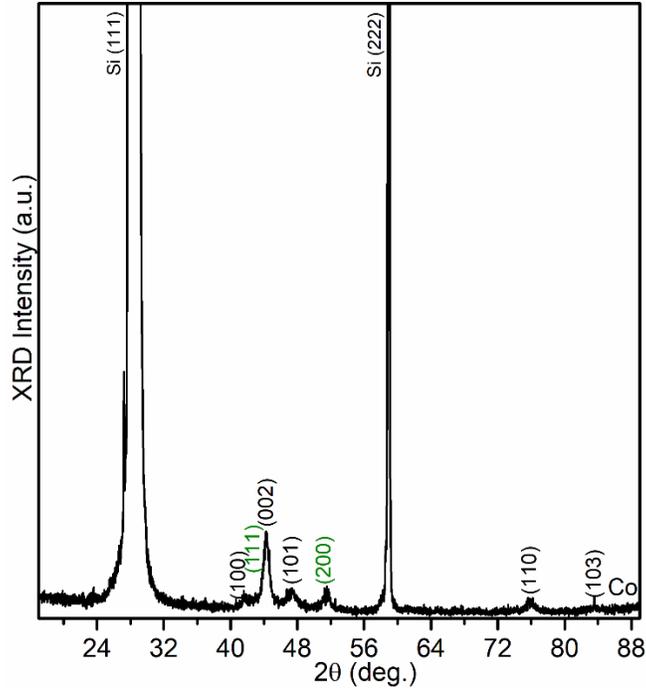
**Fig. S4** X-ray diffraction pattern of the Co nanoparticle film.

Figure S5 shows the field-dependent magnetization curves measured at 10 K and 300 K using Superconducting Quantum Information Device (SQUID). The corresponding measured coercivities are 0.18 T (Fig. S5) and 0.04 T (inset of Fig. S5), respectively.

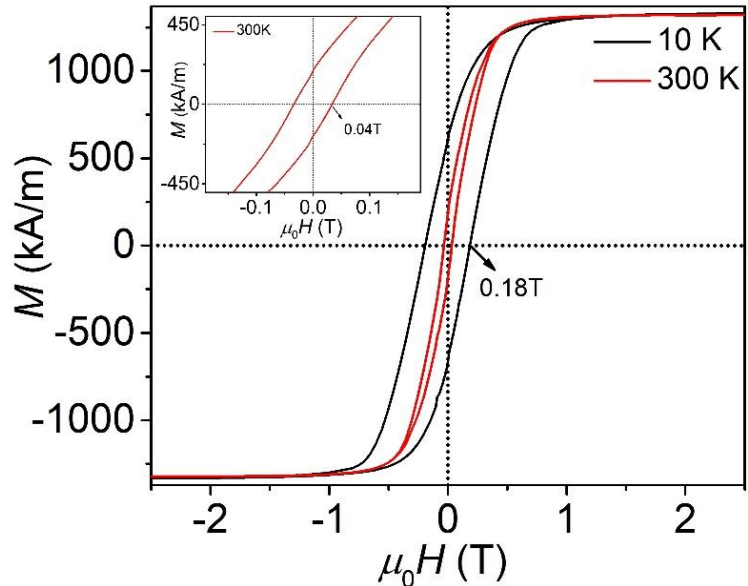

**Fig. S5.** Magnetic hysteresis loops measured at 10 K and 300 K. The expanded room-temperature loop with clear coercivity is shown as inset.

For electron-transport measurements, the film of thickness about 270 nm composed of Co nanoparticles having an average particle size of 13.7 nm, schematically shown in Fig. S3(d), was used. The conduction is through the contact point of nanoparticles as explained in Ref. 8. Therefore, a larger thickness of the film is essential to ensure the conduction through the contact points of nanoparticles (i.e., dense-packed film will have minimum voids). We also carried out micromagnetic simulations for different thicknesses of nanoparticle films and the results shows that magnetization reversal phenomena stay same giving finite skyrmion number in all cases (discussed in Fig. S17).

The THE was extracted from the Hall-effect measurements performed using physical property measurement system (PPMS). Figure S6(a) shows the experimental curves measured at 300 K used to extract the topological Halle-effect ($\rho_{THE}$) from the electron-transport and magnetic data (red curve). The extraction is based on the formula [2, 6]

$$\rho_{xy}(H) = R_0 H + R_s M + \rho_{THE}(H) \quad \quad (S1)$$

where $\rho_{xy}$ is the Hall resistivity, $R_0H$ describes the normal Hall effect, and $R_sM$ corresponds to the anomalous Hall effect. $R_0$ and $R_s$ are the ordinary and anomalous Hall coefficients, respectively. For the approximate character of this equation when applied to inhomogeneous systems, see Ref. [7].

When the magnetization approaches saturation, $\rho_{THE} = 0$ and $R_sM$ becomes constant. Therefore, $R_0$ and $R_s$ were determined by plotting $\rho_{xy}/H$ vs $M/H$ in the high field region. Using the values of $R_0$, $R_s$ and $M$, the $R_0 H$ (blue curve), $R_sM$ (black curve), and $\rho_{xy}-R_0H$ data (pink curve) were determined as a function of $H$ as shown in Fig. S6(a). The difference between the $R_sM$ and $\rho_{xy}-R_0H$ curves ideally yield the topological Hall effect contribution (green curve in Fig. S6(a)). The topological Hall effect contributions extracted from the experimental Hall data at 300 K and 10 K are shown in Fig. S6(b).

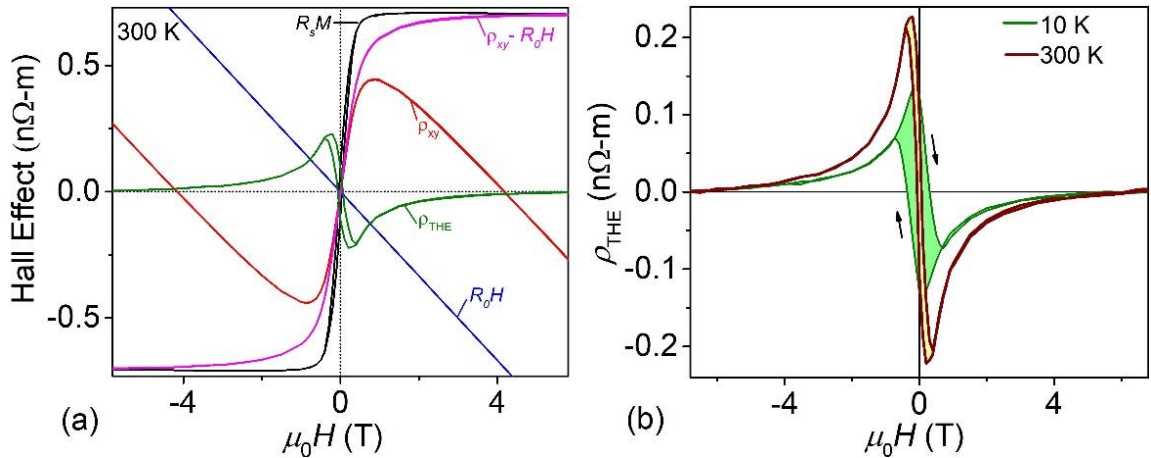

**Fig. S6.** *Hall effect and Topological Hall effect:* (a) Field dependences of Hall resistivity (red), ordinary Hall effect (blue), AHE contribution (black), Hall resistivity excluding the ordinary Hall effect (pink), and topological Hall effect (green), and (b) topological Hall effect at 300 K and 10 K.

The longitudinal resistivity, Fig. S7, shows of the Co nanoparticle thin film slightly increases with temperature, from 0.26 µΩ m at 10 K to 0.32 µΩ m at 300 K (Fig. S7). This shows that the film is metallic and that the Co nanoparticles touch each other. This metallic contact is necessary to ensure exchange coupling between the nanoparticles and a noncollinear spin structure like that in Fig. 2(c) of the main manuscript. This ensures a topological Hall effect in case of a noncoplanar spin structure and the interparticle exchange assumed in the micromagnetic simulations. Note that the Bulk resistivity of Co is lower than the Co nanoparticle films because the conduction is only through the contact points of the nanoparticles [8].

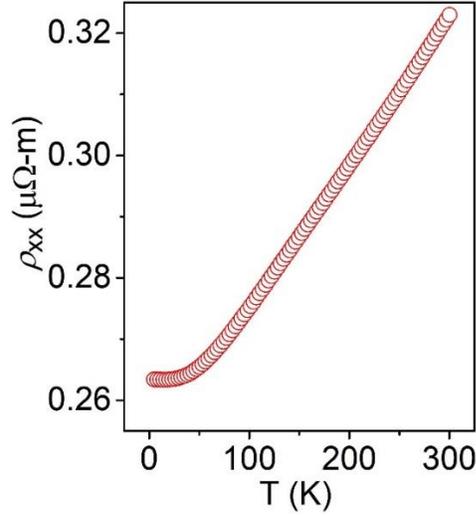

**Fig. S7.** Temperature-dependent longitudinal resistivity $\rho_{xx}$ of the Co nanoparticle film

Figure S8 provides additional information about experimental magnetic properties. The field-dependent magnetization data in Fig. S7 show how the Barkhausen jumps of Fig. 3(a) shown in the main manuscript evolve as the temperature increases.

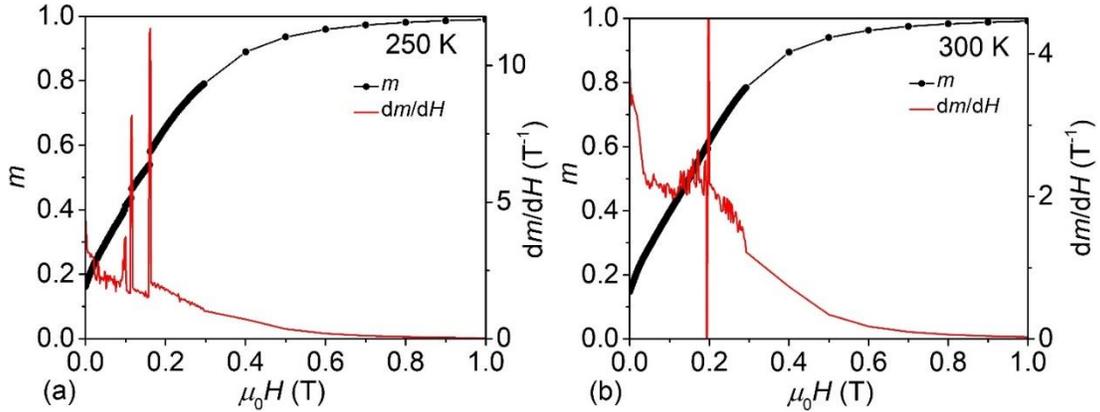

**Fig. S8.** Field-dependent normalized magnetization data measured at 250 K and 300 K and corresponding d$m$/dH curves. The data were measured during the demagnetization and $m$ is the normalized magnetization with respect to the saturation magnetization $M_s$.

Additional information about the experimental transport behavior is provided in Fig. S9, which compares Hall-effect data at different temperatures. There are no qualitative changes while the quantitative trends need further research.

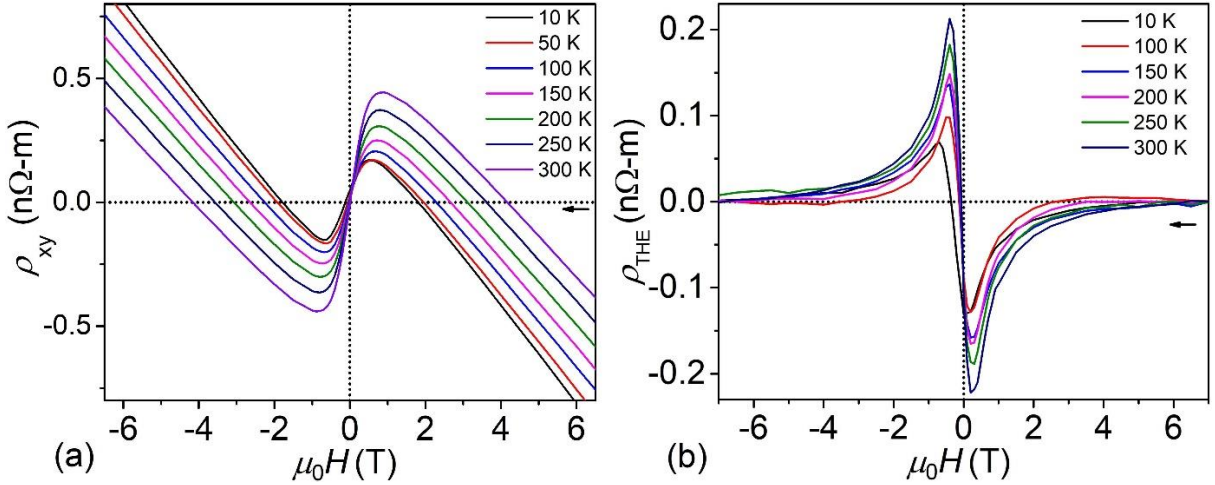

**Fig. S9.** Experimental field and temperature dependence of (a) Hall resistivity and (b) topological Hall effect.

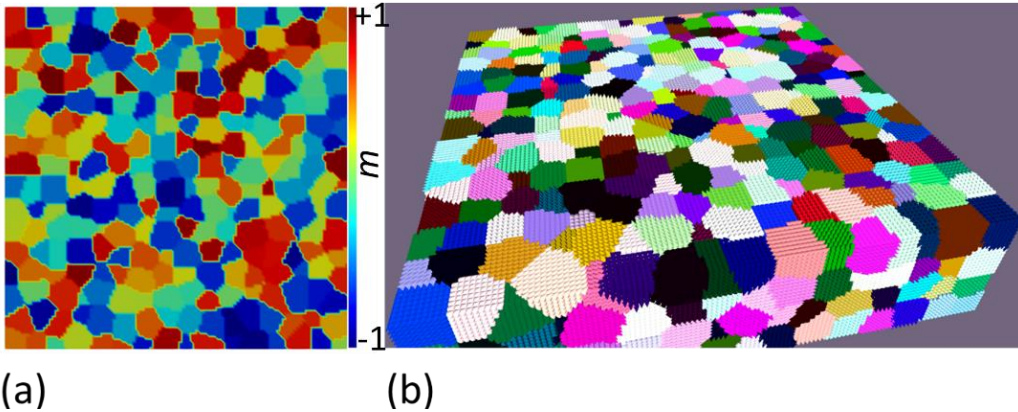

**Fig. S10.** Co nanoparticle thin film used in the micromagnetic simulations: (a) two-dimensional image showing the magnetization direction in the virgin state (initial curve at $H = 0$) and (b) three-dimensional image showing the randomness of the particles' magnetocrystalline anisotropies.

Figures S10 shows the thin-film system considered in our *ubermag* simulations. Figure S10(a) is a two-dimensional top view on the normalized magnetization $\boldsymbol{m}(r) = \boldsymbol{M}(r)/M_s$ in the virgin state, whereas Fig. S10(b) shows the random anisotropy of the nanoparticles, each color corresponding to a particle specific easy axis $\boldsymbol{n}$. The easy axes are randomly distributed, $\langle n_x \rangle = \langle n_y \rangle = \langle n_z \rangle = 0$, as imposed by the thin-film fabrication. The sizes of the regions in (a) and (b) are 240 nm × 240 nm and 240 nm × 240 nm × 60 nm, respectively. Since the atomic structure of Co is inversion-symmetric, there are no bulk Dzyaloshinskii-Moriya interactions (DMI). There are, however, small angular corrections due to random anisotropy and DMI at the surface, which ignore. Note that the Co nanoparticles exhibit nanoscale inversion symmetry, as opposed, for example, to Co/Pt bilayers.

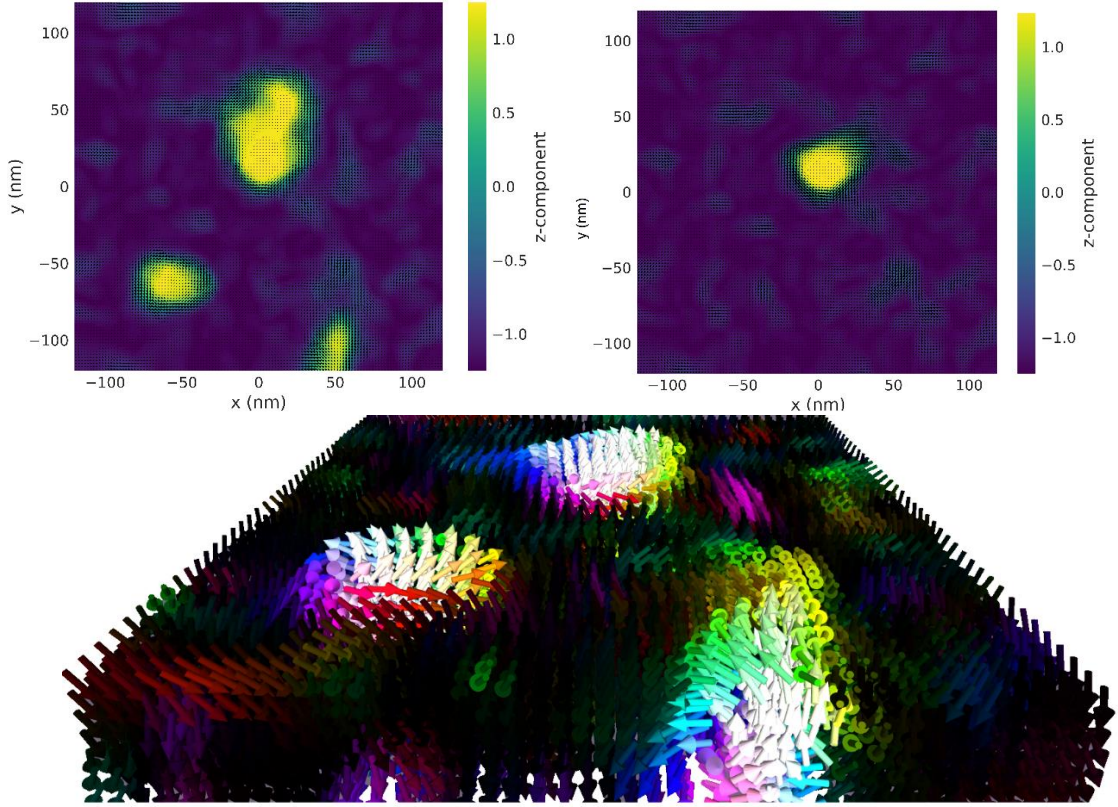

**Figure S11**: Domains separated by chiral domain walls in terms of the spin vector. Light blue represents in plane z-component of magnetization and yellow shows *m* = + 1 and blue *m* = - 1 for the z-component of magnetization. This figure also give the information related to dipolar skyrmion in micromagnetic simulations. The integration around the domain wall gives Eq. 2.

The energy functional considered during micromagnetic simulations is [9]:

$$E = \int \left\{ A\left[\nabla\left(\frac{M}{M_s}\right)\right]^2 - K_1 \frac{(n.M)^2}{M_s^2} - \mu_o M.H - \frac{\mu_o}{2} M.H_d(M) \right\} dV \quad (S2)$$

Here $M_s(r)$ is the saturation magnetization, $K_1(r)$ denotes the first uniaxial anisotropy constant, $A(r)$ is the exchange stiffness, and $n(r)$ is the unit vector of the local anisotropy direction. $H$ is the external magnetic field, and $H_d$ is the magnetostatic self-interaction field:

$$H_d(r) = \frac{1}{4\pi} \int \frac{3(r-r')(r-r')\cdot M(r') - |r-r'|^2 M(r')}{|r-r'|^5} dV' \quad (S3)$$

For complicated magnets we can use $(3rr - r^2)/r^5 = -\nabla(r/r^3)$ and $\nabla.(ab) = a \nabla.b + \nabla a.b$. It enable to write eq. S3 in terms of magnetic charge density $\rho_M = -\nabla.M$. The self-interaction then assume the form [10]:

$$E_{ms} = \frac{\mu_o}{4\pi} \int \frac{\rho_M(r)\rho_M(r')}{|r-r'|} dV' \quad (S4)$$

Since most magnets are structurally inhomogeneous, so that $\nabla.M \neq 0$ inside the magnet. This leads to relatively high energy. Which is reduced by domain wall formation. Additionally magnetic charge at the surface also leads to relatively high magnetostatic energy. Domain formation and flux closure are very good source to reduce this energy.

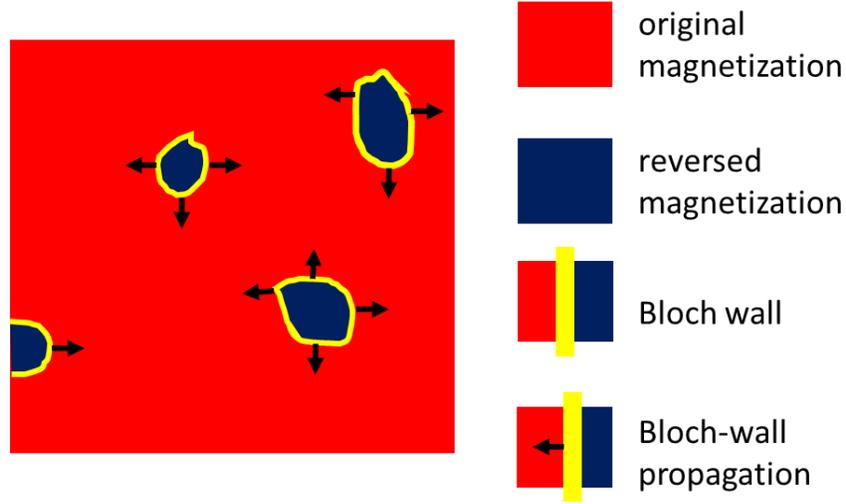

**Fig. S12.** Microstructural interpretation of domains, each creating finite magnetic flux quantum of $\pm h/e$.

Figure S12 show the formation of the magnetic domain due to the cooperative reversal of the exchanged-coupled nanomagnets. These domains are created during the reversal in order to achieve the lowest energy state. In nanostructure, the multi-domains appear due to the charge avoidance principle [55]. Our Co nanoparticle thin film is inhomogeneous, giving finite magnetic charge density $\rho_M = -\nabla \cdot \mathbf{M}$. The minimization requires that the magnetization within the film must have little divergence in order to avoid finite $\rho_M = -\nabla \cdot \mathbf{M}$. Therefore, the magnetization reversal is initiated in a small volume around inhomogeneity. During the reversal, bubble-like domain appears with core magnetization surrounded by the Bloch wall. The domain expands under the action of the reverse field, and the cylindrical Bloch domain wall subsequently forms spin texture like magnetic bubbles [12]. These type of domains are type 1 bubbles in which the core is uniformly magnetized surrounded by the Bloch type domain wall that circulate either in a clockwise or anticlockwise direction as describes in Fig. S1 of supplement A. These bubbles are mainly stabilized by dipole-dipole interactions rather than a strong Dzyaloshinskii-Moriya interaction and have the same topology of Bloch skyrmion, creating magnetic flux quantum of $\pm h/e$ and giving rise to finite THE [68].

The exchange stiffness can be estimated as a function of Curie temperature $T_c$ as [14]

$$A_{\text{ex}} = \frac{J_o S^2 \, 4.\sqrt{2}}{a} \tag{S5}$$

where $a$ is the nearest-neighbor distance between the Co magnetic atoms and is about 0.25 nm. The exchange integral $J_o$ is related to Curie temperature $T_c$ by [55]

$$J_o = \frac{3 \, k_B \, T_c}{2 \, S(S+1) z} \tag{S6}$$

Where z is nearest neighbors and in hcp its equal to 12, and $k_B$ is Boltzmann constant. So the exchange constant Eq. (S5) becomes

$$A_{\text{ex}} = \frac{k_B \, T_c \, S}{\sqrt{2} \, a \, (S+1)} \tag{S7}$$

The spin quantum number $S$ is related to saturation magnetization by $M_s = g\mu_B S/V$ [14]. $g = 2$ and $V$ is the volume. Using saturation magnetization at 10 K and 300 K in Eq. (S8) gives $A_{ex} = 10.3$ pJ/m at 10 K and 10.1 pJ/m at 300 K. This shows that the exchange constant decreases with temperature as observed in Ref. [15].

Note that our simulations on isolated nanoparticles only show curling (not shown in this study). However, the Co film investigated in the present study is composed of exchange-coupled nanoparticles and therefore the strong exchange coupling between the nanoparticles does not favor curling and causes only chiral domains via cooperative magnetization reversal. In other words, it is not possible to have curling mode as the individual particle size is very small Ref [9]. To have curled in individual nanoparticles, the necessary size of the nanoparticle should be larger so that the dipole-dipole interactions result in curling mode. In our case, the particle size is very small, so exchange interactions will keep spins parallel to each other with in the particles due to strong exchange coupling. We do not see magnetization curling but cannot exclude the presence of chiral spin [Ref. 47 of the main manuscript] at the contact point of nanoparticles, along with the presence of chiral domain which appears due to cooperative reversal. Another type of curling mode appear in the composite of hard and soft magnets [70] which is also absent in our case.

We have used low-temperature high-magnetic field scanning probe microscopy (LTHM-SPM) Attocube system to measure atomic force microscope (AFM) topography and magnetic force microscope (MFM) images. We used Co tips for measuring MFM images at 200 K using attocube and the height of the tip from the sample is maintained around 250 nm. Figure S13 shows the AFM image for the even region of Figure 4 of the main manuscript. Figure S14 shows the MFM image measured in zero magnetic field ($H = 0$) and the corresponding various regions of topography images (b-e).

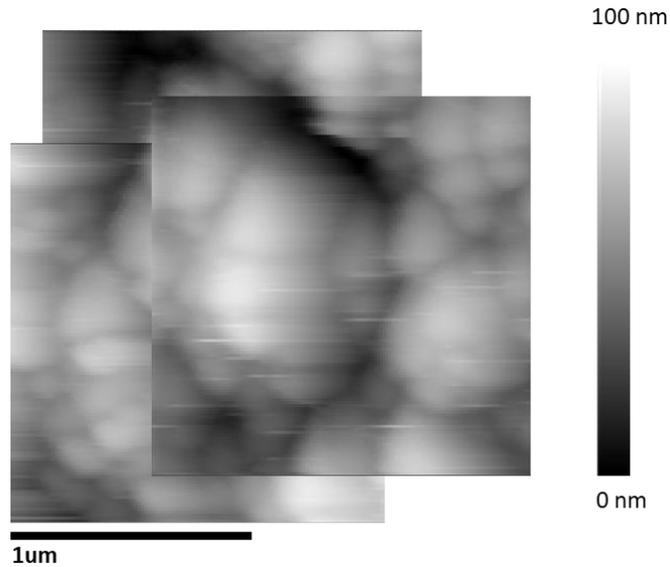

**Fig. S13**: AFM for the even region where the field gradient deeper shown in Fig. 4 (main manuscript).

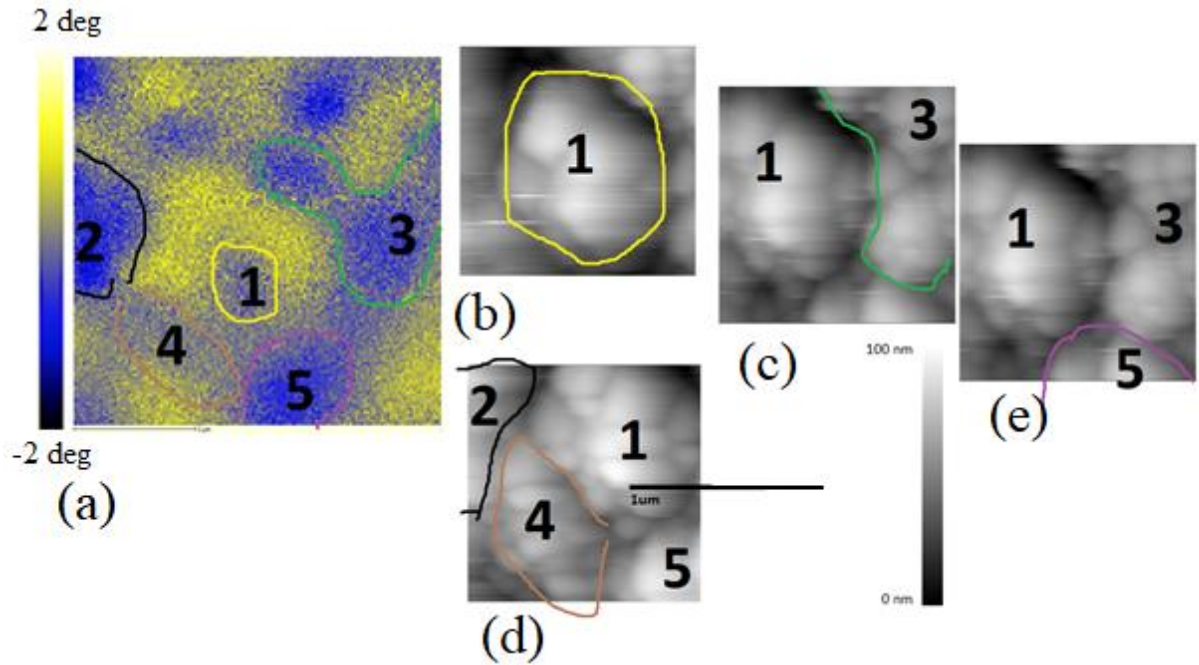

**Fig. S14**: (a) Magnetic microscopy image and (b)-(e) the topographic atomic force microscope image of Co nanoparticles film recorded at T = 200 K and H = 0 Oe of Fig 5 in the manuscript. The various regions of film 1-5 in (a) are also indicated in (b)-(e). The regions are made of clusters which are composed of nanoparticles. With the change of magnetic field these clusters change the magnetization direction and effect the neighboring clusters until the reversal expands in whole regions. As the magnetic field increases, the magnetic domains in the region (1-5) expands and coalescence of the magnetic domains occurs at the high field. Region 1-5 has an almost even surface, so the field gradient despair in these regions at high field, whereas on an uneven surface the field gradient does not disappear.

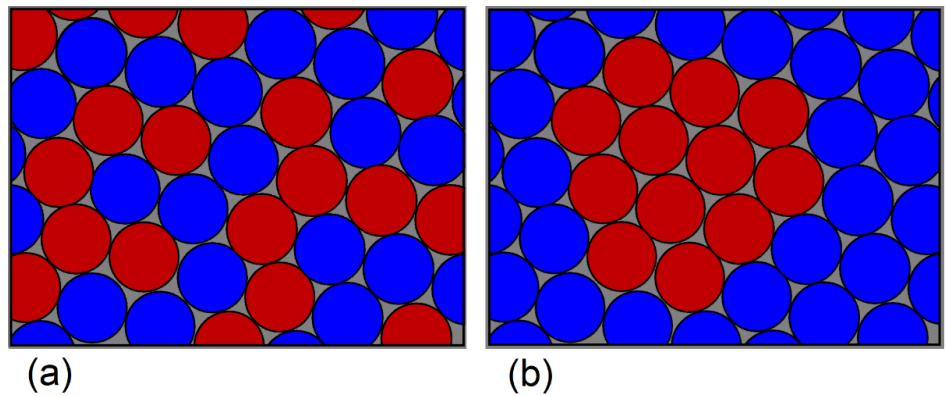

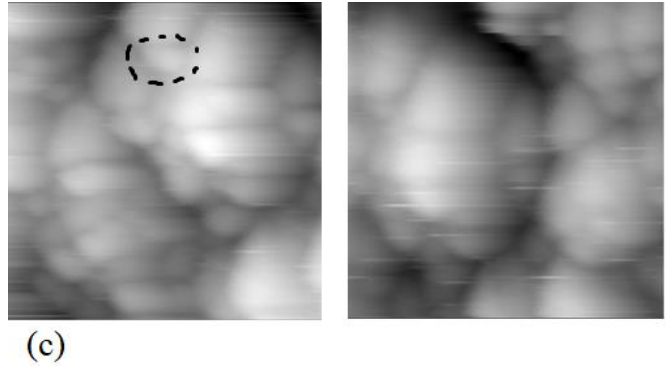

**Fig. S15.** Theoretical description of the Co nanoparticles: (a) noncooperative reversal, (b) cooperative reversal. Cooperative reversal shown in (b) creates finite magnetic flux quantum of $\pm h/e$. The cooperative reversal represents the cluster within one region. As the magnetic field increases, it expands and affect the neighboring cluster. (c) Big clusters/regions are made of small grains with visible boundaries, these small clusters are made of small nanoparticles. In cooperative reversal these grains 1$^{st}$ reverse just like in (a) and then this reversal extends to whole region.

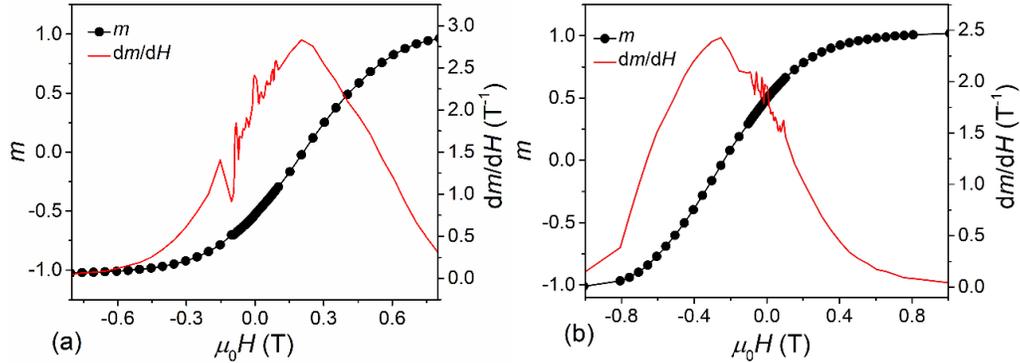

**Fig. S16.** Field-dependent normalized magnetization data measured at 10 K and corresponding d$m$/d$H$ curves for exchanged-coupled nanoparticle film of thickness of 190 nm showing Barkhausen jumps. The data were measured during the demagnetization and ***m*** is the normalized magnetization with respect to the saturation magnetization $M_s$. These data point obtained for normal step size of magnetic hysteresis. If the step size are further decreased then the height of Barkhausen jumps will increase significantly.

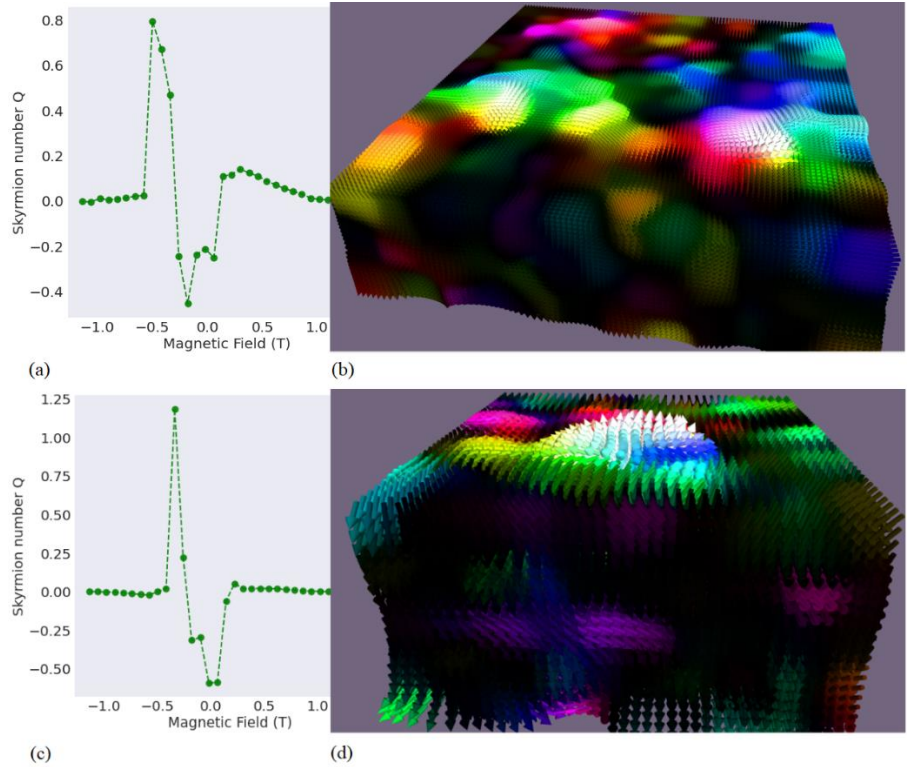

Figure S17: Micromagnetic simulations for different thicknesses. (a, b) Film dimensions 180 nm × 180 nm × 45 nm and (c, d) Film dimension of 180 nm × 180 nm × 75 nm. (a) and (c) show the change in skyrmion number as a function of the magnetic field for 45 nm and 75 nm thickness films respectively, while (b) and (d) are the corresponding spin textures, respectively when the skyrmion number is maximum. White spins (spin up) are surrounded by in-plane spins and spin downs.

**References:**
[1] J. W. Zwanziger, M. Koenig, and A. Pines, "Berry's Phase", Annu. Rev. Phys. Chern. **41**, 601-646 (1990).
[2] S. Seki and M. Mochizuki, "Skyrmions in magnetic materials", Springer International, Cham 2016.
[3] B. Ding, J. Zhang, H. Li, S. Zhang, E. Liu, G. Wu, X. Zhang, and W. Wang, Appl. Phys. Lett. **116**, 132402-1-5 (2020).
[4] W. Fenchel, Math. Ann. **101**, 238-252 (1929).
[5] B. Balasubramanian, X. Zhao, S. R. Valloppilly, S. Beniwal, R. Skomski, A. Sarella, Y. Jin, X. Li, X. Xu, H. Cao, H. Wang, A. Enders, C.-Z. Wang, K.-M. Ho, and D.J. Sellmyer, Nanoscale, **10**, 13011 (2018).
[6] A. Neubauer, C. Pfleiderer, B. Binz, A. Rosch, R. Ritz, P. G. Niklowitz, and P. Böni, Phys. Rev. Lett. **102**, 186602-1-4 (2009).
[7] R. Skomski, B. Balasubramanian, A. Ullah, C. Binek, and D. J. Sellmyer, AIP Adv, **12**, 035341 (2022).
[8] B. Balasubramanian, T. A. George, P. Manchanda, R. Pahari, A. Ullah, R. Skomski, and D. J.Sellmyer, Phys. Rev. Mater. **5**, 024402-1-9 (2021).
[9] R. Skomski, "Nanomagnetics", J. Phys.: Condens. Matter **15**, R841-896 (2003).

[10] A. A. Thiele, Bell Syst. Tech. J. **48**, 3287 (1969).

[11] J. M. D. Coey, Magnetism and Magnetic Materials (Cambridge University Press, New York, 2010), R. Skomski and J. M. D. Coey, Permanent Magnetism (Institute of Physics, Bristol, 1999).

[12] J. Jiang, D. Xiao, F. Wang, J.-H. Shin, D. Andreoli, J. Zhang, R. Xiao, Y.-F. Zhao, M. Kayyalha, L. Zhang, K. Wang, J. Zang, C. Liu, N. Samarth, M. H. W. Chan, and C.-Z. Chang, Nat. Mater. 19, 732 (2020).

[13] W. Koshibae, N. Nagaosa, New J. Phys. **18** 045007 (2016).

[14] H. Kronmüller and M. Fähnle, Micromagnetism and the Microstructure of Ferromagnetic Solids (University Press, Cambridge 2003).

[15] R. Moreno, R. F. L. Evans, S. Khmelevskyi, M. C. Munoz, R. W. Chantrell, and O. Chubykalo-Fesenko, Phys. Rev. B **94**, 104433 (2016).

[16] X. H. Zhang, T. R. Gao, L. Fang, S. Fackler, J. A. Borchers, B. J. Kirby, B. B. Maranville, S. E. Lofland, A. T. N'Diaye, E. Arenholz, A. Ullah, J. Cui, R. Skomski, I. Takeuchi J. Magn. Magn. Mater. **560**, 169627 (2022).